\newif\ifsubmode
\submodefalse

\newif\ifprintfig
\printfigtrue

\newif\ifemulate
\emulatetrue

\ifsubmode
  \documentclass[12pt,preprint]{aastex}
  \received{}
  \accepted{}
  \journalid{}{}
  \articleid{}{}
\else
   \documentclass{emulateapj}
   \submitted{{\it Accepted for publication in AJ}}
\fi
\newcommand{\kms}{\,km~s$^{-1}$}

\def\lesssim{\mathrel{\hbox{\rlap{\hbox{\lower4pt\hbox{$\sim$}}}\hbox{$<$}}}}
\def\gtrsim{\mathrel{\hbox{\rlap{\hbox{\lower4pt\hbox{$\sim$}}}\hbox{$>$}}}}

\def\spose#1{\hbox to 0pt{#1\hss}}
\def\simlt{\mathrel{\spose{\lower 3pt\hbox{$\mathchar"218$}}
     \raise 2.0pt\hbox{$\mathchar"13C$}}}
\def\simgt{\mathrel{\spose{\lower 3pt\hbox{$\mathchar"218$}}
     \raise 2.0pt\hbox{$\mathchar"13E$}}}

\slugcomment{Draft 7/14/11}

\shorttitle{Irregular kinematics of Wil 1}
\shortauthors{Willman~et~al.}

\begin{document}

\title{Willman 1 - a probable dwarf galaxy with an irregular
  kinematic distribution}

\author{Beth Willman\altaffilmark{1}, Marla Geha\altaffilmark{2}, Jay
  Strader\altaffilmark{3,4}, Louis E. Strigari\altaffilmark{5}, Joshua
  D. Simon\altaffilmark{6}, Evan Kirby\altaffilmark{7,8}, Nhung
  Ho\altaffilmark{2}, Alex Warres\altaffilmark{1}}

%Connie
%  Rockosi\altaffilmark{6}

\altaffiltext{1}{Departments of Physics and Astronomy, Haverford
  College, Haverford, PA 19041, bwillman@haverford.edu, awarres@haverford.edu}
\altaffiltext{2}{Astronomy Department, Yale University, New Haven, CT
  06520, marla.geha@yale.edu}
\altaffiltext{3}{Hubble Fellow, now Menzel Fellow}
\altaffiltext{4}{Harvard-Smithsonian CfA, Cambridge, MA
02144, jstrader@cfa.harvard.edu}
\altaffiltext{5}{Kavli Institute for Particle Astrophysics and Cosmology,  Stanford University, Stanford, CA 94305, strigari@stanford.edu}
%\altaffiltext{6}{UCO/Lick Observatory, University of California,
%    Santa Cruz, 1156 High Street, Santa Cruz, CA~95064, crockosi@ucolick.org}
\altaffiltext{6}{Observatories of the Carnegie Institution of
  Washington, Pasadena, CA 91101, jsimon@obs.carnegiescience.edu}
\altaffiltext{7}{Hubble Fellow}
\altaffiltext{8}{California Institute of Technology, Pasadena, CA
  91106, enk@astro.caltech.edu}

%%%%%%%%%%%%%%%
% Start the abstract on a fresh page
%%%%%%%%%%%%%%%

\ifsubmode\else
  \ifemulate\else
     \clearpage
  \fi
\fi

%%%%%%%%%%%%%%%
% Use a small baselineskip, unless in submission mode.
%%%%%%%%%%%%%%%

\ifsubmode\else
  \ifemulate\else
     \baselineskip=14pt
  \fi
\fi

\begin{abstract}
\renewcommand{\thefootnote}{\fnsymbol{footnote}}

We investigate the kinematic properties and stellar population of the
Galactic satellite Willman 1 (Wil 1) by combining Keck/DEIMOS
spectroscopy with KPNO mosaic camera imaging.  Wil 1, also known as
SDSS J1049+5103, is a nearby, ultra-low luminosity Milky Way
companion.  This object lies in a region of size-luminosity space
($M_V \sim -2$ mag, $d \sim $ 38 kpc, $r_{\rm half} \sim 20$ pc) also
occupied by the Galactic satellites Bo\"otes II and Segue 1 and 2, but
no other known old stellar system. We use kinematic and
color-magnitude criteria to identify 45 stars as possible members of
Wil 1.  With a systemic velocity of $v_{\rm helio} = -12.8 \pm 1.0$
\kms, Wil 1 stars have velocities similar to those of foreground Milky
Way stars.  Informed by Monte-Carlo simulations, we identify 5 of the
45 candidate member stars as likely foreground contaminants, with a
small number possibly remaining at faint apparent magnitudes.  These
contaminants could have mimicked a large velocity dispersion and
abundance spread in previous work.  We confirm a significant spread in
the abundances of the likely Wil 1 red giant branch members ([Fe/H] =
$-$1.73 $\pm$ 0.12 and $-$2.65 $\pm$ 0.12, [Ca/Fe] = $-$0.4 $\pm$ 0.18
and $+$0.13 $\pm$ 0.28).  This spread supports the scenario that Wil 1
is an ultra-low luminosity dwarf galaxy rather than a star
cluster.  Wil 1's innermost stars move with radial velocities offset by
8 \kms\ from its outer stars and have a velocity dispersion consistent
with 0 \kms\ , suggesting that Wil 1 may not be in dynamical
equilibrium. The combination of the foreground contamination and
unusual kinematic distribution make it difficult to robustly determine
the dark matter mass of Wil 1.  As a result, X-ray or gamma-ray
observations of Wil 1 that attempt to constrain models of particle
dark matter using an equilibrium mass model are strongly affected by
the systematics in the observations presented here. We conclude that,
despite the unusual features in the Wil 1 kinematic distribution,
evidence indicates that this object is, or at least once was, a dwarf
galaxy.
\end{abstract}


\section{Introduction}\label{intro_sec}
\renewcommand{\thefootnote}{\fnsymbol{footnote}} Since 2004, over a
dozen Milky Way satellites have been discovered via slight statistical
overdensities of individual stars in the Sloan Digital Sky Survey
(SDSS) catalog and confirmed by both follow-up imaging and
spectroscopy \citep[e.g.][]{willman05a,willman05b,zucker06a,zucker06b,
  belokurov06b,belokurov07a,sakamoto07,irwin07a,walsh07a,
  belokurov08a, belokurov09a}. These satellites are dominated by old
stellar populations and have absolute magnitudes of $-8 < M_V < -1$
mag. Their median $M_V$ is $\sim$ $-$4, less luminous than the median
observed for Milky Way globular clusters \citep[GCs; ][]{harrisGCcat}.
Stellar kinematics consistent with mass-to-light (M/L) ratios $>$ 100
demonstrate that most of these objects are dark matter dominated
dwarf galaxies \citep{munoz06b,simon07a,martin07a,strigari08b}.

Four of the new Milky Way companions - Willman 1, Bo\"otes II, Segue 1
and Segue 2 - contain $L \lesssim 1000 L_{\odot}$ and have been
difficult to classify.  With estimated $r_{\rm half}$ of 20 -- 40 pc,
these four objects lie in a gap between the sizes of known old stellar
populations (Milky Way GCs and dwarf spheroidals) in size-luminosity
space.  They are less luminous than all but three known objects
classified as globular clusters, providing few stars bright enough for
kinematic study
(\citealt{willman05a,belokurov07a,walsh08a,belokurov09a}). Moreover,
their proximity to the Milky Way ($d \lesssim 40$ kpc) and their
possible embedding in the Sagittarius stream (Bo\"otes II and Segue 1
and 2, \citealt{belokurov09a,NO09} - although see \citealt{law10a})
complicate the interpretation of their observed properties.

Measuring the dark mass content of satellites with $M_V > -3$ is a
critical ingredient to our understanding of the size and mass scale of
dark matter clustering, the abundance and distribution of dark matter
halos, and the extreme low mass limit of galaxy formation.
\citet{koposov07b} and \citet{walsh09a} showed that Milky Way
companions fainter than $M_V \sim -3$ could not have been discovered
at all in SDSS if they are more distant than $\sim$ 50 kpc from the
Sun.  They may thus represent the tip of an iceberg of such objects
around the Milky Way \citep[e.g.][]{tollerud08a}.  Moreover, these
objects have been shown to be excellent targets for observations
seeking the gamma-ray signature of annihilating dark matter
\citep{strigari08a,geha09a}.

Two primary lines of reasoning have been used to argue for dark matter
content in, and thus a galaxy classification for, these four extreme
objects: i) a kinematic distribution that is unbound without dark
matter \citep[Segue 1 and 2: ][]{geha09a,belokurov08a}, and/or
ii) a spectroscopically observed spread in [Fe/H], which is not
expected in purely stellar systems with a total mass as low as the
observable baryonic masses of the ultra-faint dwarfs \citep[Willman
1: ][]{martin07a}.  Thus far, the strongest evidence for a
substantial dark mass component has been provided by the line-of-sight
velocities of stars in the Segue 1 object.  \citet{geha09a} analyzed
24 member stars observed with DEIMOS to find $(M/L_V)_{\rm central} >
1000$.  Simon et al. (in preparation) confirm this result with a
larger sample of 71 member stars.  Circumstantial lines of reasoning
have also been used to argue that Bo\"otes II may contain a
substantial dark matter component \citep{walsh08a}.

The reliability of these kinematic or spectroscopic [Fe/H] analyses of
nearby dwarf galaxies hinge on having samples of member stars that
are as contamination free as possible, and a quantitative calculation
of the unavoidable contamination that may be present.  Contaminants
may be stars from the Milky Way, from unrelated
stellar streams such as Sagittarius, or from an unbound component of
the dwarf galaxy itself.  With a set of only $\sim$ 10 - 50 member
star velocities, a small number of interlopers could artificially
inflate the observed velocity dispersion, leading to an overestimate
of the mass-to-light ratio.  With only a few stars in each of these
systems bright enough for a spectroscopic [Fe/H] measurement, just one
or two interloper stars at bright apparent magnitudes could mimic an
[Fe/H] spread.  Foreground Milky Way thick disk and halo stars (at the
photometric depths reachable by spectroscopy) contaminate the
color-magnitude diagrams (CMDs) of these extreme satellites.  Segue 1
and Bo\"otes II have systemic velocities of $-$206 $\pm$ 1.3 \kms
\citep{geha09a} and $-$117 \kms \citep{koch09a}, respectively, which
are offset from the majority of thick disk and halo stars.  However,
both Willman 1 ($v_{sys} = -13.3 \pm 2.5$ \kms, \citealp{martin07a})
and Segue 2 ($v_{sys} = -40$ \kms, \citealp{belokurov09a}) have
systemic velocities that substantially overlap with the velocities of
Milky Way stars, making the identification of interlopers particularly
difficult.

Willman 1 (Wil 1; SDSS J1049+5103), located at ($\alpha_{2000},
\delta_{2000}) = (162.343^{\circ}, 51.051^{\circ}$), is an old,
metal-poor Milky Way satellite at a distance of $38\pm 7$\,kpc with
$M_V \sim -2$ mag \citep{willman05a,martin08a}.  Based on equilibrium
models of its mass, this object has been claimed to have a high dark
matter content \citep{geha09a,wolf10a}.  A high dark matter content
plus its relative proximity would make it a promising source of
gamma-rays from annihilating dark matter \citep{strigari08a}. As a
result of this prediction, several groups have attempted to
investigate the particle nature of dark matter via gamma-ray
\citep{essig09a,aliu09a} and X-ray
\citep{lowenstein10a} observations.  However, its possibly irregular
spatial distribution supports the idea that it is tidally disturbed
\citep{willman06a,martin07a} which could mean that its kinematics do
not faithfully trace its gravitational potential.  Although
\citet{martin07a} argued that Wil 1's classification of a dwarf galaxy
was supported by an observed spread in the [Fe/H] of its member stars,
such spectroscopic studies of Wil 1 suffer particularly from the
presence of interloper stars. Its systemic velocity is similar to the
velocities of Milky Way foreground stars with colors and magnitudes
consistent with Wil 1 membership.  \citet{siegel08a} showed that
contamination from Milky Way stars was a problem in the
\citet{martin07a} study and found that $2-5$ of the 7 Wil 1
spectroscopic red giant branch ``members'' were actually Milky Way
foreground stars.  These foreground stars might have generated the
apparent spread in [Fe/H].

To address the present uncertainties in the nature of the Wil 1
object, we present DEIMOS observations of 45 probable member stars.
We carefully characterize the possible contamination in this sample,
and then use it to study the abundances and kinematics of stars in Wil
1.  In \S\,\ref{sec_data} we discuss target selection and data
reduction for our DEIMOS slitmasks.  Selecting Wil 1 member stars,
including a detailed discussion of foreground contamination is
presented in \S\,\ref{sec_members}.  In \S\,\ref{sec_dwarf}, we
discuss whether Wil 1 appears to be a star cluster or a dwarf galaxy,
and then analyze Wil 1's kinematics in \S\,\ref{sec_kin}.

\section{Data}\label{sec_data}

\subsection{Photometry}\label{ssec_phot}
The data are from wide-field imaging of Wil 1 with the mosaic imager on
the 4m at Kitt Peak National Observatory (KPNO) on April 7 and 8,
2005. The reduction, photometry and calibration of these data used in
this paper is identical to that presented in \citet{willman06a}.  We
briefly describe here the reduction, photometry, and photometric
calibration of these data.  See \citet{willman06a} for more details.

Ten 600s exposures were taken through each of the SDSS $g$ and $r$
filters, with seeing varying between 1.2$''$ and 1.4$''$.  Photometry
was performed on each individual exposure and then averaged to yield a
stellar catalog with excellent photometry and star-galaxy separation
at the apparent magnitudes we are investigating.

Stellar magnitudes were photometrically calibrated with the SDSS Data
Release 4 \citep{dr4} stellar catalog. The apparent magnitudes were
then corrected for extinction using the \citet{schlegel98} dust map
values given in the SDSS catalog. The average E(g-r) along the line of
sight to Wil 1 is 0.014.  All magnitudes in this paper are
de-reddened; we use the subscript ``$_0$'' to denote that the colors
and magnitudes are de-reddened.  At magnitudes brighter than $r_0$ =
17.0, we replaced the KPNO photometry with the PSF photometry from
SDSS Data Release 7 \citep{dr7}.

\begin{figure}[t!]
\plotone{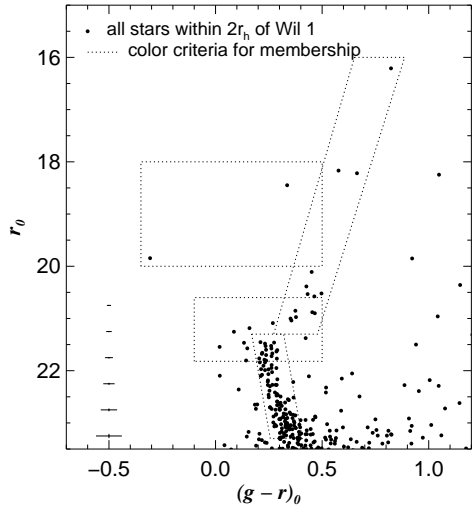}
\caption{Dereddened color-magnitude diagram of all stars within two
  elliptical half-light radii of the center of Willman~1 from KPNO
  $g$- and $r$-band photometry.  We used (position angle, ellipticity,
  $r_{\rm half}$) = (77,0.47,2.3$'$) from \citet{martin08a} to calculate
  half-light distances.  The region inside the dotted boxes is the
  location of our highest priority spectroscopic selection criteria,
  hereafter referred to as the color criteria used to identify stars
  possibly belonging to Wil 1.  The sizes of color and magnitude
  uncertainties are shown by the crosses on the left of the CMD.
  \label{fig_cmd}}
\end{figure}

Figure~\ref{fig_cmd} shows the CMD of all point sources within two
elliptical half-light radii of the center of Wil 1, using the
structural parameters derived by \citet{martin08b}.  The points in
this CMD delineate the main sequence of Wil 1, with candidate red
giant branch (RGB) stars visible.  Dotted boxes outline the liberal
color-magnitude selection that we will use in the rest of this paper
as the color-magnitude requirements for possible Wil 1 membership.

We applied the color-magnitude selection shown in Figure 1 to this
photometric catalog to calculate a revised center of Willman 1.  We
began with the center, based on the much shallower SDSS dataset,
calculated by \citet{martin08b}, and then iteratively calculated the
average position of stars within 2 arcminutes of the center until we
converged on ($\alpha_{2000}, \delta_{2000}$) = (162.3397,51.0508).  We
will use this center for the rest of the paper.

%%%%%%%%%%%%%%%%%%%%%%%%%%%%%%%%%%
% Figure:  CMD
%%%%%%%%%%%%%%%%%%%%%%%%%%%%%%%%%%

\subsection{Spectroscopic Target Selection}\label{ssec_targets}

Stars in Wil 1 were targeted for spectroscopy using the photometric
catalog described in the previous section. We set the target
priorities to preferentially observe stars with a high likelihood of
being Wil 1 members based on their color, magnitude and spatial
position.  First priority was given to stars that (1) spatially
overlap the main body of Wil 1 and (2) reside within regions of the
color-magnitude diagram that are consistent with the Main Sequence
(MS) and turnoff, horizontal branch, and red giant branch of an old
stellar population at the distance of Wil 1.  These color-magnitude
criteria are shown by the dotted lines overplotted on
Figure~\ref{fig_cmd}.  We chose to implement liberal, rectangular
color-magnitude criteria to include Wil 1 member stars with a range of
possible [Fe/H] and ages in our spectroscopic sample.  Second priority
was given to stars occupying a similar color-magnitude region,
independent of spatial location.  All remaining stars were assigned
third priority.  Within each of these three tiers, stars were further
prioritized by their apparent magnitude, with the brightest stars
receiving highest priority.  An average of 100 slitlets were placed on
each mask (see Table~1).

\subsection{Spectroscopy and Data Reduction}\label{ssec_redux}

Four multislit masks were observed for Willman~1 using the Keck~II
10-m telescope and the DEIMOS spectrograph \citep{faber03a}.  Three
masks were observed on the nights of November 20--22, 2006, the fourth
was observed on March 20, 2007. Exposure times, mask positions and
additional observing details are given in Table~1.  The masks were
observed with the 1200~line~mm$^{-1}$ grating covering a wavelength
region $6400-9100\mbox{\AA}$.  The spatial scale is $0.12''$~per
pixel, the spectral dispersion of this setup is $0.33\mbox{\AA}$, and
the resulting spectral resolution is $1.37\mbox{\AA}$ (FWHM). Slitlets
were $0.7''$ wide.  The seeing conditions during both runs were on
average $\sim 0.75''$.  Despite the similar observing conditions, few
spectra were usable from the fourth mask because the targeted stars
were fainter.  The minimum slit length was $4''$ to allow adequate sky
subtraction; the minimum spatial separation between slit ends was
$0.4''$ (three pixels).

Spectra were reduced using a modified version of the {\tt spec2d}
software pipeline (version~1.1.4) developed by the DEEP2 team at the
University of California-Berkeley for that survey. A detailed
description of the two-dimensional reductions can be found in
\citet{simon07a}.  The final one-dimensional spectra are rebinned into
logarithmically spaced wavelength bins with 15\,\kms\ per pixel.

\subsection{Radial Velocities and Error Estimates}\label{ssec_rvel}

We measure radial velocities and estimate velocity errors using the
method detailed in \citet{simon07a}.  We refer the reader to this
paper for a description of the method and only highlight the important
steps below.

Radial velocities were measured by cross-correlating the observed
science spectra with a series of high signal-to-noise stellar
templates.  The templates were observed with Keck/DEIMOS using the
same setup as described in \S~\ref{ssec_redux} and cover a wide
range of stellar types (F8 to M8 giants, subgiants and dwarf stars)
and metallicities ([Fe/H] = $-2.12$ to $+0.11$).  We calculate and
apply a telluric correction to each science spectrum by cross
correlating a hot stellar template with the night sky absorption lines
following the method in \citet{sohn07}.  The telluric correction
accounts for the velocity error due to mis-centering the star within
the $0.7''$ slit caused by small mask rotations or astrometric errors.
We apply both a telluric and heliocentric correction to all velocities
presented in this paper.

It is crucial to accurately assess our velocity errors because
the internal velocity dispersion of Willman~1 is expected to be
comparable to the DEIMOS velocity errors associated with individual
measurements.  We determine the random component of our velocity
errors using a Monte-Carlo bootstrap method.  Noise is added to each
pixel in the one-dimensional science spectrum.  We then recalculate the
velocity and telluric correction for 1000 noise realizations.  Error
bars are defined as the square root of the variance in the recovered
mean velocity in the Monte-Carlo simulations.  The systematic
contribution to the velocity error was determined by \citet{simon07a}
to be 2.2\kms\ based on repeated independent measurements of
individual stars, and has been confirmed by a larger sample of repeated
measurements.  While we did not place stars on multiple masks in
the Willman~1 dataset, the systematic error contribution is expected
to be constant as this is the same spectrograph setup and velocity
cross-correlation routines are identical.  We add the random and
systematic errors in quadrature to arrive at the final velocity error
for each science measurement. 

The fitted velocities were visually inspected to ensure the
reliability of the fit.  Radial velocities for 111 stars of the 423
objects targeted passed this visual inspection. For the rest of this
paper, we only consider the 97 of those 111 stars that have spectra
with $S/N > 2$, and that have velocity uncertainties of
less than 10 \kms.  The median velocity uncertainty of these 97 stars
is 3.5\kms.  The median velocity uncertainty in the subsample of those
stars most likely to be Wil 1 members is 4.7\kms, because the faintest
of these 97 stars are dominated by main sequence stars belonging to
Wil 1.  All velocity histograms shown in this paper thus have a
binsize of 4.7\kms.

\subsection{Comparing relative and deriving absolute abundances of individual stars}\label{ssec_feh}

In \S3.3 and \S3.4 we will use the relative stellar abundances
imprinted on these spectra to identify likely bright member stars, and
to compare with stars observed by the SEGUE survey.  In \S4.1 we will
then calculate the absolute [Fe/H] abundances of three bright member
stars.  We describe here the techniques used to measure those relative
and absolute abundances.

To assess relative stellar abundances, we use the reduced equivalent
width (W$'$) of the Ca II lines.  We utilize two functional forms of
W$'$, one which has been calibrated for low metallicity stars
\citep{starkenburg10} and the standard \citet{rutledge97}
calibration. For both, we measure the CaT lines at $8498, 8542$, and
$8662\mbox{\AA}$ using the continuum and line definitions described in
\citet{rutledge97} to calculate $\Sigma {\rm Ca}$.  We determine the
uncertainty on $\Sigma {\rm Ca}$ with the Monte Carlo method described
above.  Added in quadrature to the Monte Carlo uncertainties is a
systematic uncertainty of $0.25$\,\mbox{\AA}, determined using the
same method described in \citet{simon07a}, but using a larger sample
of repeated measurements.

Throughout the paper, we primarily report the value of the
\citet{rutledge97} definition of W$'$ such that $\Sigma {\rm Ca} =
0.5\rm{EW}_{8498\AA}+ \rm{EW}_{8542\AA} + 0.6\rm{EW}_{8662 \AA}$ and
W$' = \Sigma {\rm Ca} - 0.64(V_{\rm HB} - V)$.  $V$ is the $V$-band
magnitude of each RGB star and $V_{\rm HB}$ is the magnitude of the
horizontal branch.  To obtain $V$, we convert our SDSS $g$- and
$r$-band magnitudes into $V$-band using the photometric
transformations given in Tables 1 and 2 of \citet{blanton07}.  The
apparent $V$ magnitude of the one spectroscopically confirmed star in
the flat part of Wil 1's horizontal branch is 18.45 mag.  Although
this approach minimizes assumptions, using a single star to determine
$V_{HB}$ introduces a possible offset into our W$'$ calculations; if
the star is variable then its current apparent magnitude may not equal
its average value.  However, \citet{siegel08a} found no RR Lyrae stars
in Wil 1 in their time-series imaging of the object.  Regardless, a
small shift the $V_{\rm HB}$ we use to calculate W$'$ would not affect
any conclusion of this paper.

We also use the metallicity calibration detailed in
\citet{starkenburg10} to calculate [Fe/H] from W$'$, where W$'$ is
defined as $0.43( V- V_{\rm HB} ) + \Sigma {\rm Ca_{(2+3)}} - 2
(\Sigma {\rm Ca}_{(2+3)}^{-1.5} )+ 0.034(V- V_{\rm HB})$, and $\Sigma
{\rm Ca_{(2+3)} }= \rm{EW}_{8542\AA} + \rm{EW}_{8662 \AA}$. This study
excluded the Ca II line at $8498 \AA$ because it is the weakest of the
three lines and has been shown to contribute more to the relative
uncertainty than the two stronger lines \citep {armandroff91}. The
CaT-[Fe/H] relation presented in \citet{starkenburg10} was calibrated
for RGB stars in the metallicity range $-4.0 < [Fe/H] \sim -0.5$ and
which lie within $-3 < (V-V_{\rm{HB}}) < 0$ or $ -3 < \rm(M_V) < 0.8$.

To derive the absolute abundances of bright member stars, we use the
spectral synthesis method of \citet{kirby08b} (KGS08).  This method
relies on comparing an observed medium-resolution spectrum with a grid
of synthetic spectra covering a range of effective temperature
($T_{\rm eff}$), surface gravity (log $g$), and composition. All of
our spectra with high enough S/N for this technique are bright enough
to be in the SDSS database, so we determine V and I for each star by
transforming the SDSS $gri$ magnitudes with the relationship derived
by \citet{jordi06}.  KGS08 found that uncertainty in the measured
colors, ages, and alpha-abundances of stars does not substantially
affect the estimated $T_{\rm eff}$ and log $g$. The best-matching
composition is then found by minimizing residuals between the observed
spectrum and a smoothed synthetic spectrum matched to the DEIMOS
spectral resolution. To derive error bars on the best-matching [Fe/H],
we calculate $\chi^2$ contours for every star by allowing both [Fe/H]
and $T_{\rm eff}$ to vary.  We separately varied $T_{\rm eff}$ by $\pm
125$ K and $250$ K and log $g$ by $\pm 0.3$ dex and $0.6$ dex.  The
surface gravity makes almost no difference in [Fe/H] because there are
no visible ionized Fe lines in red giants in the observed spectral
range.  Changing $T_{\rm eff}$ by $\pm 125$ K yields a $\delta$[Fe/H]
of 0.13 dex and by an unrealistically large $\pm 250$ K yields a
$\delta$[Fe/H] of 0.26 dex for each star we study in this paper.
KGS08 found that their technique measured [Fe/H] with 0.25 (0.5) dex
accuracy on spectra of Galactic globular cluster stars with S/N $\sim$
20 ($\sim$ 10) $\AA^{-1}$.  The success of this method has been
confirmed by a comparison with high-resolution abundances of over 100
stars in GCs, the halo field, dwarf galaxies, and of six RGB
stars with $-$3.3 $<$ [Fe/H] $<$ $-$2.3 in the ultra-faint dSphs
\citep{kirby08a,kirby10a}.

\section{A Spectroscopic Sample of Wil 1 Stars}\label{sec_members}

To study the stellar population and kinematic properties of Wil 1, we
need to identify a sample of member stars with minimal contamination
from interlopers.  Wil 1 lies at relatively high Galactic latitude
$(l,b)$ = $(158.6^{\circ}, 56.8^{\circ})$. However, its systemic
velocity is $\sim -12.3\pm2.5$ \kms\ \citep{martin07a} and the median
velocity of all Milky Way stars in the direction of Wil 1 is $-15.0$
\kms\ using the Besancon Galaxy model \citep{robin03}. In addition to
overlapping in velocity, stars in Wil 1 have colors and magnitudes
very similar to the colors and magnitudes of stars in the Milky Way's
thick disk and halo.

%\footnote{http://bison.obs-besancon.fr/modele}

In this section, we use color, magnitude, and velocity to select an
initial sample of 45 Wil 1 candidate member stars.  We then derive the
predicted contamination from Milky Way foreground stars within this
member sample.  Finally, we identify 5 stars that are probable
interlopers and present the properties of the remaining 40 probable
Wil 1 member stars.

\subsection{A Color-Magnitude-Velocity Sample of Wil 1 Candidate
  Stars}\label{ssec_members}

The first required criterion for Wil 1 membership is having a color
and magnitude consistent with the stellar population of Wil 1.  We use
a loose color-magnitude (CM) selection, as shown by boxes overplotted
with dotted lines on Figure~\ref{fig_cmd}.  We used this loose cut,
rather than proximity to a fiducial cluster sequence, to avoid making
a priori assumptions about the stellar population of Wil 1.  58 of the
97 stars in our spectroscopic catalog satisfy these CM criteria.  

\begin{figure}
\plotone{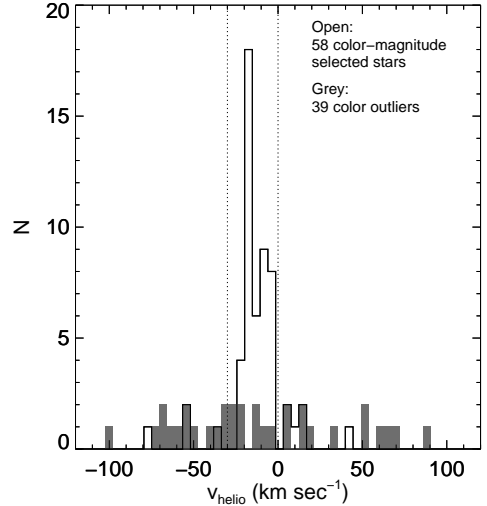}
 \caption{Velocity distributions of: the 58 stars that satisfy our Wil
 1 color-magnitude selection criteria (open) and the 39 stars that do
not satisfy these criteria (grey filled).  The dotted lines show the
velocity range of $-30 < v_{\rm helio} < 0$ \kms used to select
Wil 1 member stars.  Binsize is 4.7 \kms, the median velocity error of
the 58 stars passing the color-magnitude criteria for membership.
\label{fig_vhist}}
\end{figure}

The second required criterion for Wil 1 membership is a velocity
consistent with belonging to Wil 1.  Figure~\ref{fig_vhist} shows the
heliocentric velocity histogram of these 58 stars, along with the
velocity histogram of the 39 stars that do not satisfy the
color-magnitude selection for Wil 1 membership.  The velocity
distribution of the 58 CM selected stars is strongly peaked, with 45
stars between $-30$ and 0 \kms.  We identify these 45
color-magnitude-velocity (CM-V) selected stars as likely Wil 1
members.  This does not necessarily mean that none of the 13 CM
selected stars with outlying velocities are physically associated with
Wil 1.  However, the spatial distribution of those 13 stars at
outlying velocities is not clustered around the Wil 1 center.

We present in Table 2 the equatorial coordinates, $r$ magnitudes,
$g-r$ colors, heliocentric velocities, and spectral S/N of the 45 CM-V
selected Wil 1 member stars.  We also include the CaT W$'$ (and
uncertainty) for the 15 possible red giant branch, as calculated in
\S~\ref{ssec_feh}.  Table 3 contains the same data (but not W$'$) for
the 52 non-member stars.

%%%%%%%%%%%%%%%%%%%%%%%%%%%%%%%%%%
% Figure:  velocity histogram
%%%%%%%%%%%%%%%%%%%%%%%%%%%%%%%%%%

\subsection{Predicting the Number of Interlopers in the
  Color-Magnitude-Velocity Sample}\label{ssec_fg}

\begin{figure}[t!]
\plotone{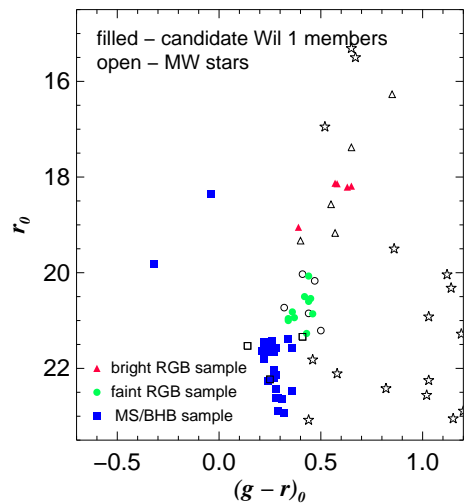}
 \caption{Color-magnitude diagram of the 97 stars with DEIMOS/Keck
 velocities. Open symbols show Milky Way stars.  Filled symbols show
 probable Wil 1 member stars, as selected by color-magnitude and
 velocity ($-30 < v < 0$ \kms) criteria.  Triangles, circles, and
 squares highlight stars belonging to the bRGB, fRGB, and MS/BHB
 sub-samples used to characterize foreground contamination.  5-point
 stars show those stars that did not satisfy the initial
 color-magnitude cut for membership.
\label{fig_colorcmd}}
\end{figure}

Figure~\ref{fig_colorcmd} shows a CMD of the stars in our
spectroscopic catalog.  Filled symbols represent the 45 candidate Wil
1 members selected in \S~\ref{ssec_members}, and open symbols
represent the 52 foreground Milky Way stars.  The number of open
symbols overlapping with the filled symbols shows that shows that 40\%
of stars with colors and magnitudes consistent with the red giant
branch of Wil 1 are foreground stars belonging to the Milky Way.
These foreground stars were only identified because their
line-of-sight velocities were different than those of Wil 1 stars.
The median velocity of Milky Way stars passing the CM criterion for
membership is $-35.7$ \kms (based on the Besancon Galaxy model), with
16\% of these having $-30 < v_{\rm los} < 0$ \kms.  How many Milky Way
interlopers remain in the CM-V sample of 45 candidate Wil 1 members?

%%%%%%%%%%%%%%%%%%%%%%%%%%%%%%%%%%
% Figure:  Velocity selected CMD
%%%%%%%%%%%%%%%%%%%%%%%%%%%%%%%%%%

We simulate the number of interloper stars expected among the 45
candidate members using the Besancon Galaxy
model.  Because photometric studies suggest the presence of tidal features
around Wil 1 \citep{willman06a,martin07a}, we first predict the number
of Milky Way contaminant stars without assuming that all CM selected
stars outside the Wil 1 velocity peak belong to the Milky Way.  We
instead use the Besancon model to predict the absolute number density
of Milky Way stars satisfying the color-magnitude-velocity criteria
for candidate members.  The predicted number of contaminant stars thus
rests on the assumptions that the velocity distribution of Besancon
model stars and the absolute numbers of stars in the Besancon model
are correct.  We later verify that this yields a reasonable
prediction.

The primary ingredients in our calculation are:

\begin{enumerate}

\item $n_{\rm fg,vel}$, the projected number density of Milky Way
  stars in the Besancon model satisfying the CM-V criteria for Wil 1
  membership. We calculated $n_{\rm fg,vel}$ and its dispersion in
  1000 small fields randomly placed in a 1 square degree Besancon
  simulation centered on the position of Wil 1.  To do this, we
  shuffled the RAs and Decs of Besancon model stars before selecting
  each random field. The random fields each had an area approximately
  equal to that of our spectroscopic survey footprint.  Because the CM
  cuts applied to our data were liberal, we simply used the model
  CFHT-Megacam $g$ and $r$ magnitudes as a proxy for the observed SDSS
  $g$ and $r$ magnitudes. We convolved 4.7 \kms\ measurement
  uncertainties, the median for the 45 candidate members, to the model
  velocities of each Besancon star.  The average number of possible
  interlopers in the CM-V sample within a given area of sky, $A$, is
  then $N_{\rm cont,vel} = A * n_{fg,vel}$.

\item $f_{\rm targ}$, the fraction of stars in our photometric catalog
  satisfying the CM criteria for Wil 1 membership that also end up in
  our spectroscopic catalog of 97 stars. Not all stars satisfying the
  CM criteria for membership were targeted, and not all targeted stars
  had spectra with high enough S/N to be in the final spectroscopic
  catalog.  To derive the true number of Milky Way interlopers,
  $N_{\rm cont,vel}$ needs to be multiplied by $f_{\rm targ}$. Because
  both the density of stars and fraction of sky covered by
  observations decreases with increasing Wil 1 distance, $f_{\rm
  targ}$ is a function of distance from the center of Wil 1.  We thus
  calculate $f_{\rm targ}$ in each of three distance ranges: 0 -- 2
  r$_{\rm half}$, 2 -- 4 r$_{\rm half}$ and 4 -- 6 r$_{\rm half}$ .

\end{enumerate}

The contamination in the Wil 1 candidate member sample is expected to
be primarily composed of stars belonging to the Milky Way's thick disk
and halo.  Because the relative number of thick disk and halo stars
varies across the CMD (thick disk stars dominate at brighter apparent
magnitudes, halo stars at fainter) and because these two galaxy
components have different velocity distributions, $n_{\rm fg, vel}$ is
a function of apparent magnitude. Because stars were prioritized by
apparent magnitude in target selection, $f_{\rm targ}$ is also a
function of apparent magnitude.  We therefore separately estimate the
foreground contamination in three subsets of the Wil 1 population: i)
the five relatively bright red giant branch (bRGB; $r_0$ $<$ 19.5
mag), ii) the 10 faint red giant branch (fRGB; $19.5 < r_0 < 21.3$
mag); and iii) the 30 main sequence (MS) and blue horizontal branch
(BHB) stars.

\begin{figure*}[t!]
\epsscale{0.9}
\plottwo{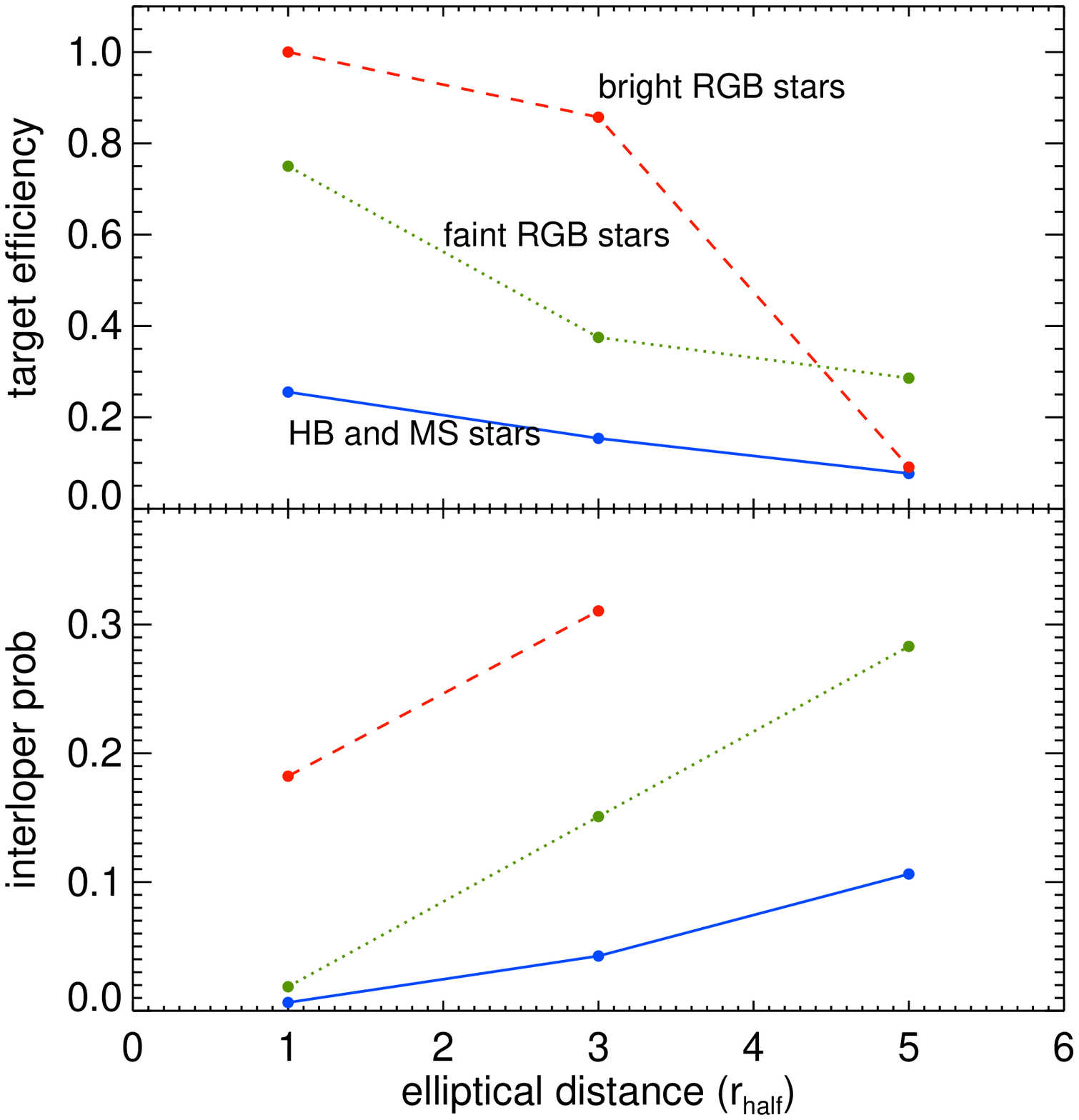}{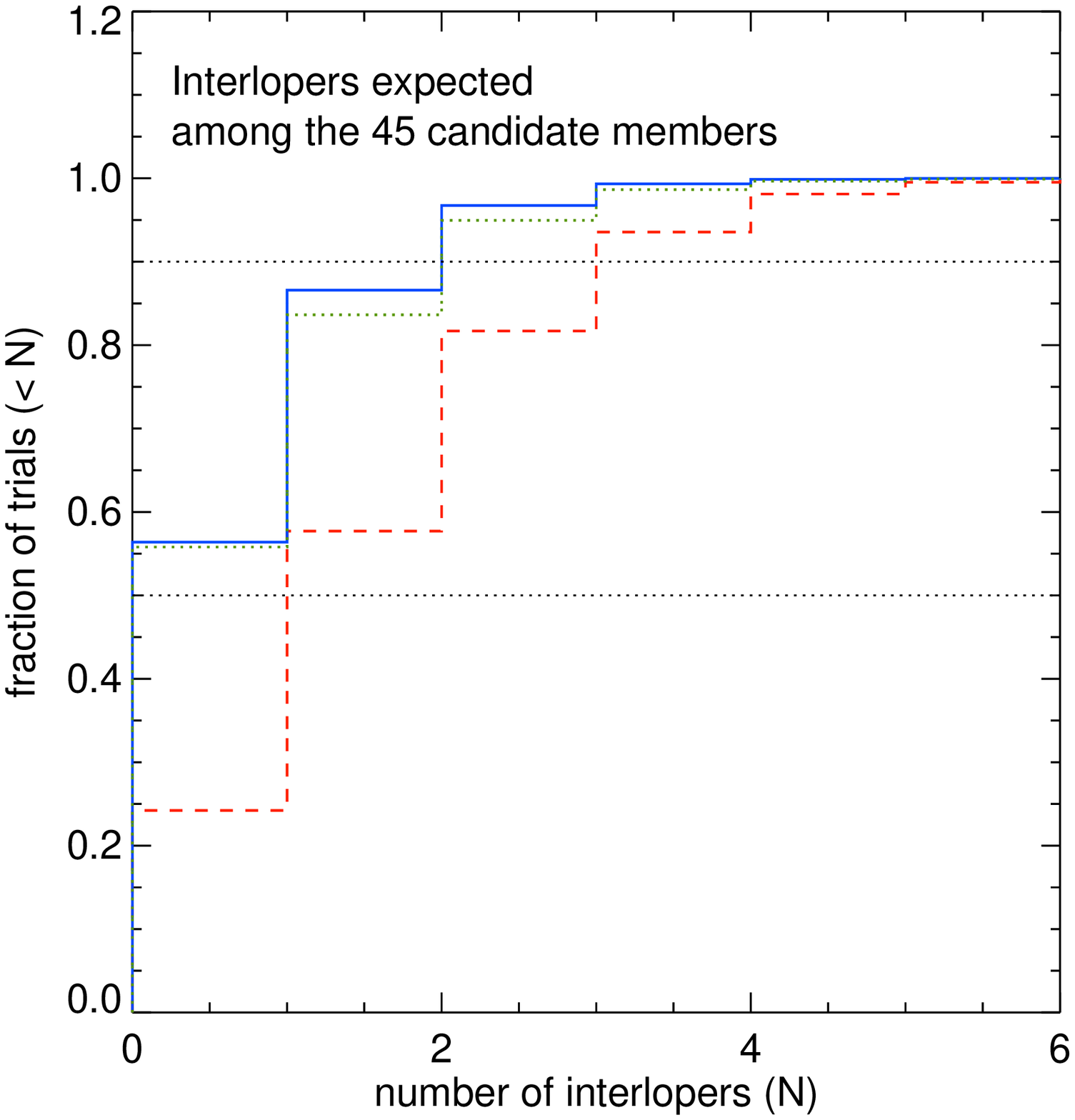}
\caption{Left panel, top: The fraction of all stars, as a function of
  elliptical distance from the Wil 1 center, that pass color-magnitude
  criteria for Wil 1 membership and also end up in our final
  spectroscopic sample with well-measured velocities.  This target
  efficiency is shown for three different subpopulations of Wil 1
  stars - bRGB candidates, fRGB candidates, and BHB/MS candidates.  Left
  panel, bottom: The probability that any individual star is an
  interloper, as a function of distance.  Right panel: The predicted
  number of Milky Way interlopers in the sample of 45 candidate
  members in $10^5$ Poisson realizations of the foreground. Color
  coding is the same as the left panel. 90\% of all trials had 7 or
  fewer interlopers (3 bRGB, 2 fRGB, and 2 MS/BHB).}
\label{fig_fg}
\end{figure*}

The left panel of Figure~\ref{fig_fg} shows $f_{\rm targ}$ as a
function of distance from Wil 1 for the bRGB, fRGB, and MS/BHB
subsamples.  The overall target efficiency at all distances is much
lower than one, because many stars at the faint end of our acceptable
magnitude range ($r_0$ = 23) for target selection were not measured
with high enough S/N for a robust velocity to be extracted.  Many
stars remain to be observed in the center of Wil 1 if such faint
magnitudes can be reached.  The target efficiency for fRGB stars
remains high out to large distances from Wil 1, even though our
fractional spatial coverage beyond 3$r_{\rm half}$ is quite low.  This
is because our mask placement included two out of the six total stars
between 4 and 6$r_{\rm half}$ that are consistent with the fRGB of Wil
1. Although target efficiency is low at large $d$, only 4\% of stars
are expected to lie beyond 3 elliptical half-light radii of a system
well-described by an exponential profile (see also
\S~\ref{ssec_distrib}).

The average expected number of contaminant stars in each of the bRGB, fRGB, and
MS/BHB CM-V subsamples in each annulus is:

\begin{equation}
N_{\rm cont,vel} = f_{\rm targ} \times A_{\rm annulus} \times n_{\rm fg,vel},
\end{equation}

and the average dispersion in this number is:

\begin{equation}
\sigma_{\rm cont,vel} = f_{\rm targ} \times A_{\rm annulus} \times \sigma_{\rm fg,vel},
\end{equation}

where $A_{\rm annulus}$ is the area of each elliptical annulus, and
$f_{\rm targ}$, $n_{\rm fg,vel}$ and $\sigma_{\rm fg,vel}$ are
calculated separately for each of the bRGB, fRGB, and MS/BHB
subpopulations.  We predict a fraction of one interloper star within
each of the three subpopulations within each of the three annuli.  The
probability that any individual star is an interloper increases with
distance and is shown in the bottom left panel of Figure~\ref{fig_fg}.

We sum the predicted average numbers (and dispersions, in quadrature)
of interlopers in each annulus, to derive the average total numbers
(and dispersions) of interloper stars expected in each of the bRGB,
fRGB, and MS/BHB subsamples and find 1.5 $\pm$ 1.0, 0.7 $\pm$ 0.5, and
0.6 $\pm$ 0.3 stars, respectively.  Although the number of stars in
our sample increases at fainter apparent magnitudes, the number of
expected contaminants is low in the 30 possible main-sequence members.
This small predicted contamination results from the low overall target
efficiency combined with the large number of Wil 1 stars in its main
sequence relative to its bRGB and fRGB.  To convert these fractional
numbers of stars into a physical number of stars that may be in our
sample, we used the IDL function POIDEV to generate $10^5$ Poisson
random deviates with the predicted mean and dispersion in the number
of interloper stars.  The result of this simulation is shown in the
right panel of Figure~\ref{fig_fg}.  50\% of all trials contained one
or zero predicted interlopers among the entire candidate sample of 45
stars - one red giant branch star.  90\% of all trials had 7 or fewer
interlopers (3 bRGB, 2 fRGB, and 2 MS/BHB).

%%%%%%%%%%%%%%%%%%%%%%%%%%%%%%%%%%
% Figure:  FOREGROUND CALCULATION
%%%%%%%%%%%%%%%%%%%%%%%%%%%%%%%%%%

To sanity check this prediction, we also use the above method to
predict the number of Milky Way stars expected in our dataset outside
the velocity cut for membership in our original CM-selected sample.
At 50\% (90\%) confidence, the number of stars with $v < -30$
\kms\ or $v > 0$ \kms\ we predict to be in the bRGB, fRGB, and MS/HB
regions of our spectroscopic dataset are $\leq$ 5, 4, and 4 (8, 7, and
7), respectively.  Figure~\ref{fig_colorcmd} shows that there are
actually 5, 5, and 3 stars with $v < -30$ \kms\ or $v > 0$ \kms\ in the
RGB, fRGB, and MS/HB regions of our spectroscopic dataset.  Our
technique thus accurately predicts the number of stars in our
spectroscopic catalog at outlying velocities.

We also use a simpler technique that does not rely on the absolute
calibration of the Besancon model to predict the number of interloper
stars among the 45 candidate members.  This ``scaled histogram''
approach instead relies on assuming that the 13 bRGB, fRGB, and
MS/HB-colored stars stars at outlying velocities are all Milky Way
foreground stars.  Treating each of these three regions of the CMD
separately, we determine the relative numbers of Milky Way stars in
the Besancon model with velocities inside and outside the velocity cut
for membership and apply this scaling factor to the numbers of stars
we observe outside the velocity cut for membership.  We use these
fractional numbers of stars to generate $10^5$ Poisson realizations of
the predicted number of bRGB, fRGB, and MS/HB interlopers and find that
50\% (90\%) of trials contain $\leq$ 1, 1, and 0 (3, 2, and 2)
interlopers respectively.  This technique thus yields very similar
results as the full simulations, predicting a median of only 2
interlopers, with fewer than or equal to 7 at 90\% confidence.

We summarize these predictions for interlopers among the 45
candidate members in Table 4.

\subsection{Identifying Interloper Stars in Wil 1 Sample}\label{ssec_IDcontam}

The calculations in \S~\ref{ssec_fg} revealed that we expect $1-7$
Milky Way interlopers among the 45 Wil 1 candidate stars identified in
\S3.1. When broken down by subsets of stellar population, we predict
$1-3$ interlopers among the 5 bRGB candidates, $0-2$ interlopers among
the 10 fRGB candidates, and $0-2$ interlopers among the 30 MS/HB
candidates.  We now attempt to identify these interlopers. Because the
fractional contamination of the candidate MS/HB Wil 1 members is small
compared to that of RGB stars, and because we do not have a
reasonable spectroscopic [Fe/H] indicator for the MS/HB candidates, we
only look for the $1-5$ interlopers with RGB magnitudes and
colors.

We use CaT reduced equivalent width, W$'$ (calculated using the
Rutledge definition described in \S2.5), to flag possible interlopers.
All recent spectroscopic studies of Milky Way dwarf galaxies use some
metallicity indicator to select member stars
\citep[e.g.][]{walker09b}.  This selection means that the abundance
spread we will infer for Wil 1 in \S4.1 is necessarily a lower limit.
We choose not to use velocity and position information to perform a
likelihood analysis for member selection.  We will show evidence that
Wil 1 is both spatially and kinematically disturbed in \S~4.2 and 5.2.
We therefore do not want to assert that Wil 1 stars follow an
exponential spatial distribution and Gaussian velocity distribution
when discriminating member stars from interlopers.

\begin{figure*}
\epsscale{0.5}
\plotone{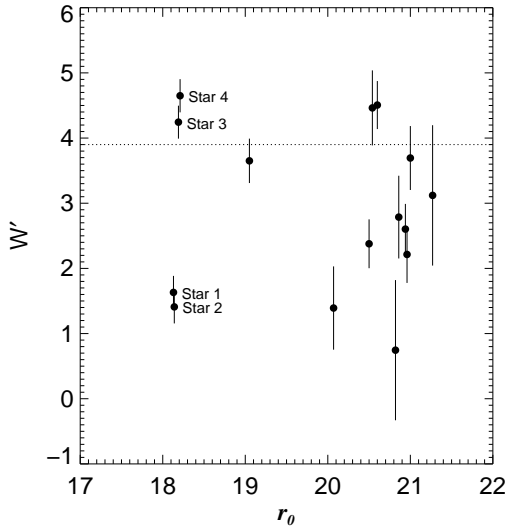}
\caption{The reduced equivalent widths (Rutledge definition) of the
  calcium triplet feature measured for the 15 candidate Wil 1 member
  stars brighter than the base of the red giant branch.  This figure
  shows a large spread in W$'$, with a gap between the more metal-rich
  and more metal-poor stars at the brightest magnitudes.  Overplotted
  is a line at 3.9 \AA, our subjective W$'$ cut between probable
  members and possible interloper stars.}

\label{fig_CaT}
\end{figure*}

Figure~\ref{fig_CaT} shows W$'$ of the 15 candidate Wil 1 RGB stars as
a function of $r_0$ magnitude. This figure shows a large spread of
W$'$, with a big gap between the more metal-poor and more metal-rich
stars at bright magnitudes.  Given the gap at bright magnitudes, we
hypothesize that the 4 higher W$'$ (more metal-rich) stars on this
figure are possible foreground stars. The dotted line at W$'$ = 3.9
\AA\ in Figure~\ref{fig_CaT} shows our adopted separation between
possible foreground dwarfs and Wil 1 members.  Using this W$'$ cut,
there are 4 likely foreground RGB stars (Stars 3, 4, 10, and 11 in
Table 2).

The fainter stars in Figure~\ref{fig_CaT} do not show the same
bi-modal distribution of W$'$ as the brighter stars.  However, we
adopt a cut at W$'$ = 3.9 \AA\ as a reasonable way to flag possible
interlopers because i. in \S3.4 we will provide additional support
for an interloper classification of Stars 3 and 4 and ii. we will show
that this specific W$'$ cut impacts neither our conclusions about the
classification of Wil 1 nor our conclusions about the kinematic
properties of Wil 1.

%%%%%%%%%%%%%%%%%%%%%%%%%%%%%%%%%%
% Figure: [Fe/H]
%%%%%%%%%%%%%%%%%%%%%%%%%%%%%%%%%%

We test this W$'$ cut by considering in detail the two
brightest stars flagged as interlopers. One of these stars, Star 4,
has a high-resolution HET spectrum from \citet{siegel08a}. They could
not obtain a consistent solution for Fe I and Fe II abundances under
the assumption that this star was a giant. They argue it is likely to
be a foreground dwarf, consistent with our classification.  Star 3 has
W$'$ and SDSS colors very similar to that of Star 4 (see Tables 2
and 5). We thus consider it likely that Star 3 is also a foreground
star, although this statement is not conclusive.  \citet{martin07a}
also classified Star 3 as an interloper.

In addition to the four relatively high W$'$ stars, we classify
the star with $r \sim 19$ and W$' \sim 3.2$ (Star 6 in Table 2)
as an interloper.  Figure~\ref{fig_colorcmd} shows that Star 6 ($g-r =
0.39$) lies blueward of the main bRGB locus in $g-r$, despite its
intermediate W$'$.  For Star 6 to reasonably be part of Wil 1's
stellar population, it would need to be quite metal-poor (inconsistent
with its intermediate W$'$) to explain its blue color.  Star 6 is
thus a likely non-member, bringing the total number of stars flagged
as contaminants on the RGB to 5 out of 15, which is at the upper edge
of our contamination estimates.

To facilitate the reader reaching her own conclusions about the
classification of the five brightest Wil 1 RGB candidate members, the
SDSS de-reddened $ugriz$ magnitudes of these stars are compiled in
Table 5.  To facilitate the reader's comparison with earlier
spectroscopic studies of Wil 1, we also include the star IDs used by
\citet{siegel08a} and the velocities used by \citet{martin07a} in this
table.

While we cannot rule out the possibility that there are relatively
metal-rich stars in Wil 1, we have outlined a reasonable approach to
flagging likely RGB non-members.  We exclude these five stars from our
primary analysis in the remainder of this paper.

\subsection{A close look at the membership probability of Stars 1 and 2}

There is an abundance of circumstantial evidence for a member
classification for both Stars 1 and 2.  In their high-resolution HET
spectroscopic study, \citet{siegel08a} concluded that Star 2 (unlike
Star 4) is a Wil 1 member giant.  Star 1's photometric and
spectroscopic properties are very similar to that of Star 2, providing
some evidence that they both are true member stars.  The positions and
velocities of Stars 1 and 2 are both typical of those of the
other 43 candidate member stars: Star 1 is only 0.6 projected
half-light radii from the center of Wil 1 and has a heliocentric
velocity of $-5.4 \pm 2.2$ \kms. 13\% of the candidate members have
more positive velocities than Star 1.  Star 2 is 2.1 projected
half-light radii from the center of Wil 1 and has a heliocentric
velocity of $ -18.5 \pm 2.2$ \kms. 27\% of the candidate members lie
at greater distance and 22\% at more negative velocities.  (We discuss
this large velocity spread and the correlation between distance and
velocity in \S5).  We have also shown it to be unlikely that we have
missed any foreground contaminants in the part of Wil 1's RGB that
includes Stars 1 and 2.

We proceed with caution and now independently test the hypothesis that
Star 1 or Star 2 may be a Milky Way halo star. We will show that Star
1's [Fe/H] is $-1.73 \pm 0.12$, similar to the peak of the halo's
metallicity distribution function \citep{ryan91a}. We will then argue
that Wil 1 is a dwarf galaxy, or remnant thereof, based the [Fe/H]
spread between Stars 1 and 2.  This classification of Wil 1 thus
hinges on the membership of both Stars 1 and 2.

We use the SEGUE survey database \citep{yanny09a} to investigate
whether Stars 1 and 2 are similar to Milky Way field stars with
similar colors, magnitudes and velocities.  We do the same for Stars 3
and 4, which we believe to be interlopers.  We search for stars
similar to Stars 1 and 2 by selecting SEGUE spectra of stars with the
following restrictive set of properties: (i) $0.51 < (g-r)_0 < 0.55$,
(ii) $1.29 < (u-g)_0 < 1.37$, (iii) $18 < g_0 < 19$, and (iv) a radial
velocity between $-30$ and 0.  We search for stars similar to Stars 3
and 4 by selecting all SEGUE spectra of stars with: (i) $0.63 <
(g-r)_0 < 0.67$, (ii) $1.67 < (u-g)_0 < 1.74$, (iii) $18 < g_0 < 19$,
and (iv) a radial velocity between $-30$ and 0.  We chose these
photometric cuts based on the SDSS DR7 magnitudes of these stars (in
Table 5).  The reported SEGUE Stellar Parameter Pipeline quantities
for the stars satisfying the Star 1/2-like (Star 3/4-like) criteria
have a median [Fe/H] = $-0.65$ ($-0.7$) and log $g = 4.42$ ($4.47$),
indicating that the samples are dominated by thick disk dwarfs, as
expected for the region of the color-magnitude diagram we are
studying.

\begin{figure*}
\epsscale{0.8}
%\plotone{fg_CaThisto.forpaper.eps}
\plotone{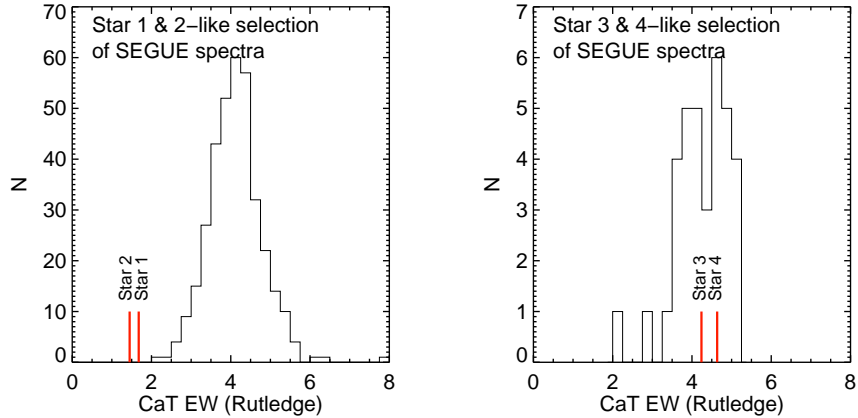}
\caption{Histograms of the CaT EW measured from stellar spectra
  obtained from the SEGUE database.  SEGUE stars were selected as
  those with color, apparent magnitude, and radial velocity extremely
  similar to those of Stars 1 \& 2 or Stars 3 \& 4 (criteria listed in
  \S3.4).  Stars 1 and 2 are both outliers from the 354 SEGUE stars
  with similar properties, showing that they are each inconsistent
  with being a foreground star.  Even without the other evidence that
  Star 1 is a Wil 1 member star, this comparison implies the chance of
  Star 1 belonging to the Milky Way foreground to be less than
  1/350. Conversely, Stars 3 and 4 are both very similar to the 35
  SEGUE stars with similar properties, supporting our hypothesis that
  they are foreground stars.}

\label{fig_CaT_fg}
\end{figure*}

We downloaded the spectra of the 355 unique SEGUE stars satisfying
Star 1/2-like criteria and of the 35 unique SEGUE stars satisfying the
Star 3/4-like criteria.  One spectrum of the 355 was excluded from
analysis because it appeared to be flawed based on a visual
inspection.  We measured the CaT EW of each SEGUE spectrum in the same
manner as for the Wil 1 stars, using the radial velocities given in
the SEGUE database (median error = 2.5 km s$^{-1}$).  The distribution
of CaT EW for the SEGUE stars is reasonably well-described as a
Gaussian with a mean of 4.1 \AA\ and a standard deviation of 0.6 \AA.
The histograms are shown in Figure~\ref{fig_CaT_fg}.

%%%%%%%%%%%%%%%%%%%%%%%%%%%%%%%%%%
% Figure: CaT foreground histo
%%%%%%%%%%%%%%%%%%%%%%%%%%%%%%%%%%
 
No star in the Star 1/2-like SEGUE sample has a CaT EW $<$ 2.2 \AA.
The Rutledge CaT EW of Star 1 is $1.68\pm0.3 \AA$ and Star 2 is
$1.45\pm0.3 \AA$ - outliers from the SEGUE stars.  The CaT EW of Star
1 is 2$\sigma$ lower than the lowest of 354 stars in the SEGUE sample,
showing that it is a true outlier from Milky Way foreground
stars. Conversely, the stars in the Star 3/4-like SEGUE sample are
very similar to Stars 3 and 4, which have CaT EW of 4.24 $\pm$ 0.3
\AA\ and 4.64 $\pm$ 0.3 \AA\, respectively.

The spectral abundances of Stars 1 and 2 provide additional
evidence against the hypothesis that either of them is a field Milky
Way halo star.  Using the spectral synthesis method of KGS08,
described in \S 2.5, [Fe/H]$_{Star1}$ = $-1.73 \pm 0.12$,
[Fe/H]$_{Star2}$ = $-2.65 \pm 0.12$, [Ca/Fe]$_{Star1}$ = $-0.4 \pm
0.18$, and [Ca/Fe]$_{Star2}$ = $+0.13 \pm 0.28$.  Although its [Fe/H]
is very similar to the [Fe/H] of field halo stars, the [Ca/Fe] of Star
1 is much lower than that of typical halo stars \citep{venn04}.  The
[Fe/H] of Star 2 is also much lower than that of typical halo stars.
In \S~\ref{ssec_spread} we will discuss the abundances of Stars 1 and
2 in more detail.

\begin{figure*}
\epsscale{0.6}
%\plotone{isos.final.eps}
\plotone{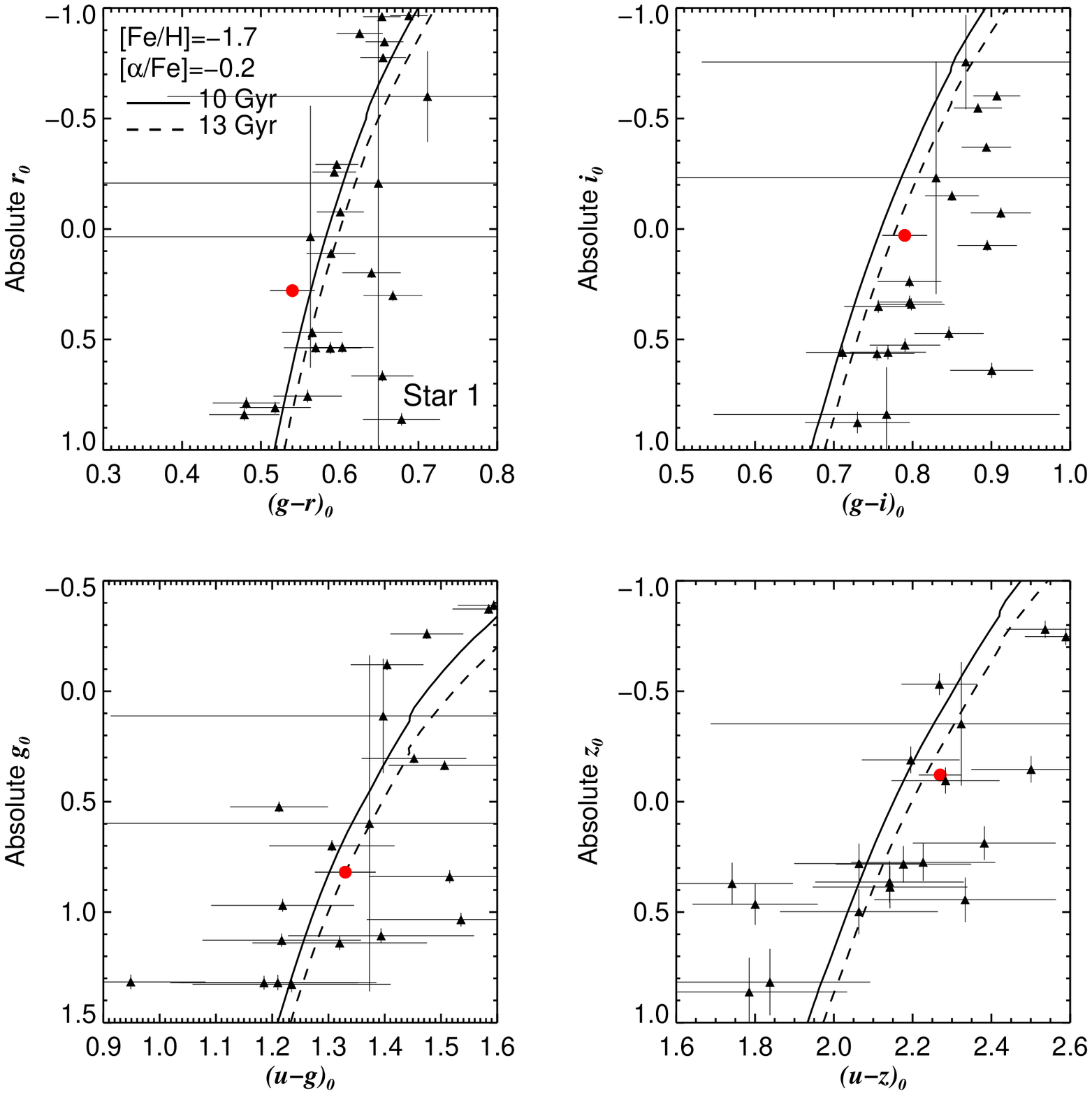}
\caption{Color-magnitude diagrams showing Star 1 (red dot) compared to
  Draco stars (black triangles) and \citet{dotter08} isochrones.
  Draco stars with $-2.0 <$ [Fe/H] $< -1.7$, as measured by
  \citet{kirby10a} are included in the figure.  All magnitudes are
  de-reddened SDSS DR7 magnitudes, converted to absolute
  magnitudes. 1$\sigma$ color-magnitude measurement uncertainties are
  shown for each star.  The isochrone is [Fe/H] = $-1.7$, [$\alpha$/Fe]
  = -0.2 because it was the available isochrone that was closest to
  the measured [Fe/H]$_{Star1}$ = $-1.73 \pm 0.12$, [Ca/Fe]$_{Star1}$
  = $-0.4 \pm 0.18$.  We have offset the model $z$-isochrones by -0.06
  mag and the model $i$-isochrones by +0.03 mag, to put them in the
  DR7 SDSS system (A. Dotter, private communication).}

\label{fig_isos1}
\end{figure*}

\begin{figure*}
\epsscale{0.6}
%\plotone{isos.final.eps}
\plotone{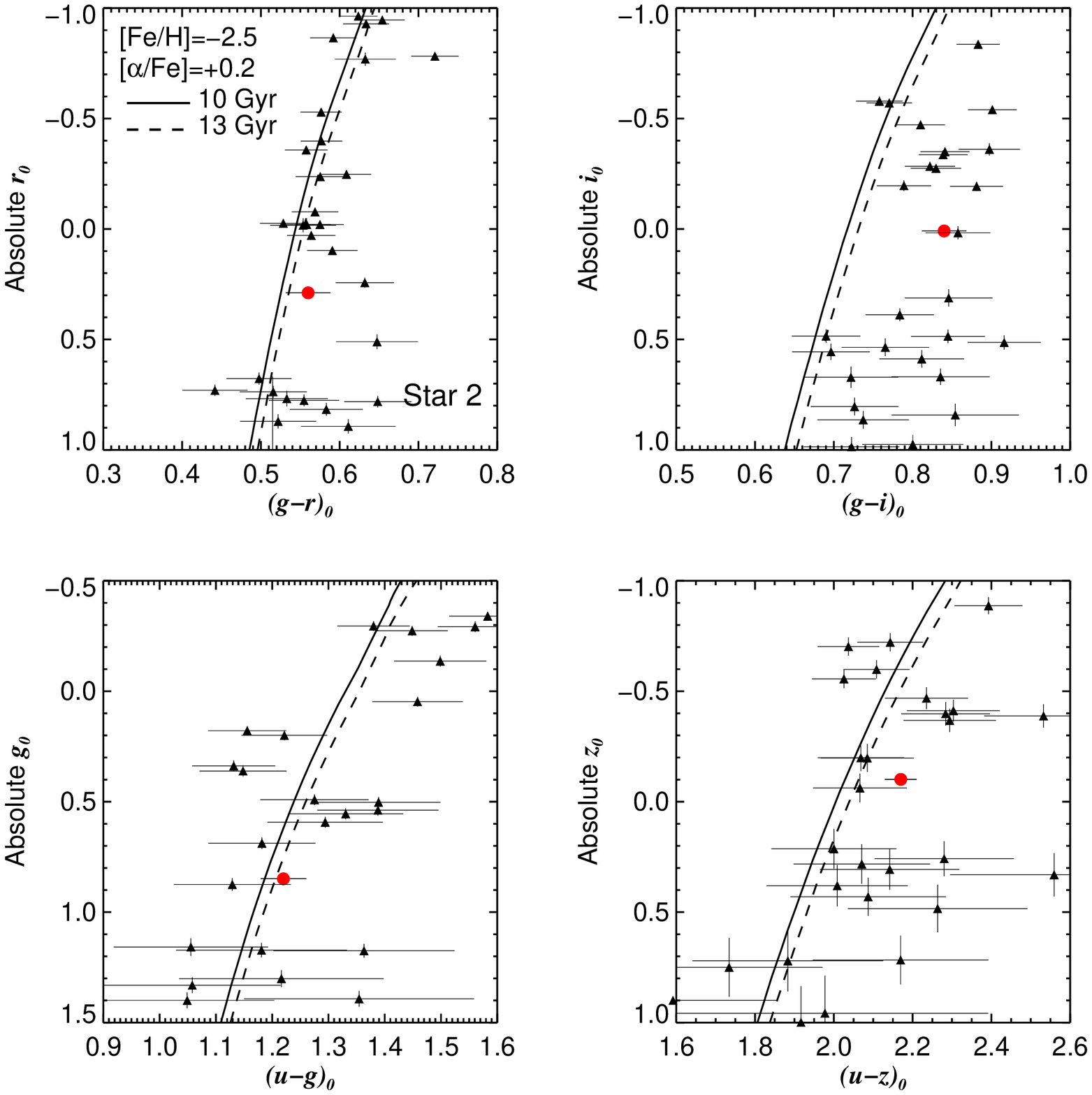}
\caption{The same as Figure 7, but for Star 2 (red dot) plus Draco
  stars (black triangles) with [Fe/H] $< -2.3$ (includes stars with
  [Fe/H] as low as $-3.4$, \citealp{kirby10a}).  The isochrone is
  [Fe/H] = $-2.5$, [$\alpha$/Fe] = +0.2 because it was the available
  isochrone that was closest to the measured [Fe/H]$_{Star1}$ = $-2.65
  \pm 0.12$, [Ca/Fe]$_{Star2}$ = $+0.13 \pm 0.28$.}

\label{fig_isos2}
\end{figure*}

Now that we have shown that neither Star 1 nor Star 2 is likely to be
a foreground star, we compare their colors and absolute magnitudes
with isochrones from the \citet{dotter08} library, and with the colors
and absolute magnitudes of stars in the Draco dSph in
Figures~\ref{fig_isos1} and \ref{fig_isos2}.  We do this comparison to
determine whether the very similar 5-color SDSS photometry of Stars 1
and 2 (Table 5) can be consistent with the hypothesis that they are at
the same distance but have very different metallicities.  The Draco
stars (black triangles) overplotted with Star 1 are those with $-2.0
<$ [Fe/H] $< -1.7$, as spectroscopically measured by
\citet{kirby10a}.  The Draco stars overplotted with Star 2 are those
with $-3.4 <$ [Fe/H] $< -2.3$, as spectroscopically measured by
\citet{kirby10a}.

These figures show that, for the most metal-poor stars, the models are
largely consistent with the data in $g-r$ and $u-g$, but are too blue in
$g-i$ and $u-z$.  This discrepancy between models and data is
qualitatively similar to that found by \citet{an09a}, who showed that
isochrones in the SDSS photometric system are systematically bluer
than the colors of main sequence stars in relatively metal-poor
globular clusters in $g-i$, $g-z$, $u-g$.

These figures also show that stars belonging to Draco can have quite
different spectroscopic metallicities, yet also have very similar
colors in the SDSS filter set.  Based on this comparison, we conclude
that the similar 5-color SDSS photometry of Stars 1 and 2 is
consistent with our hypothesis that they both belong to Wil 1 and have
[Fe/H] that differ by 1 dex.

%%%%%%%%%%%%%%%%%%%%%%%%%%%%%%%%%%
% Figure: CaT foreground histo
%%%%%%%%%%%%%%%%%%%%%%%%%%%%%%%%%%

Although our comparisons demonstrate that the simplest explanation for
Stars 1 and 2 is that they belong to Wil 1 rather than the field halo,
we cannot use this analysis to completely rule out the possibility
that one is a halo star.  For example, we could be unlucky and Wil 1
may lie in a direction and at a distance that has an excess of
stellar halo structure with abundances atypical relative to other
lines-of-sight.  However, given all of the evidence, we conclude that
Stars 1 and 2 are both Wil 1 members.

%%%%%%%%%%%%%%%%%%%%%%%%%%%%%%%%%%
% Figure: lots of stuff
%%%%%%%%%%%%%%%%%%%%%%%%%%%%%%%%%%

\subsection{A comparison with other methods of identifying interlopers}

Other groups have used different approaches to identify likely Milky
Way interlopers among spectroscopic samples of dwarf galaxy stars.
The expectation maximization approach, tailored by
\citet[][]{walker09b} for the study of Milky Way dwarf galaxies,
combines the line-of-sight velocity, Mg equivalent width, and
projected position of each star in spectroscopic samples of Fornax,
Carina, Sextans, and Sculptor stars to derive a probability that each
star is a dwarf galaxy member rather than a contaminant star. While
this approach is a powerful tool for studying the more luminous Milky
Way galaxies, it is not easily applicable to the extremely low
luminosity satellites. First, unlike the \citet{walker09b} datasets,
the spectroscopic samples of the extreme satellites tend to include
stars with a wide range of surface gravities. Second, we want to drop
the assumptions that the velocity distributions of stars in the
extreme satellites are well described by Gaussians and that stars
follow undisturbed exponential profiles out to large distances.

Another method frequently used to identify interlopers is the
equivalent width of the Na I line at 8190.5 \AA. This line lies in our
spectral range, is sensitive to both surface gravity and temperature,
and has previously been used to discriminate between red giant branch
stars and red main sequence stars. However, \citet{gilbert06}
demonstrate that while this spectral feature is very effective for
dwarf/giant discrimination for $V-I > 2.5$, it is not effective for
stars bluer than $V-I = 2.0$. The red giant members of Wil 1 in our
sample have $0.4 < V-I < 1.0$, so we cannot use the Na I line to
distinguish them from foreground dwarfs. It is possible that the use
of this diagnostic by \citet{martin07a} in their spectroscopic study
of Wil 1 led to the inclusion of interlopers in their Wil 1
sample. For example, they classify our Star 4 as a member giant even
though \citet{siegel08a} show from a high-resolution spectrum that it
is very likely to be a foreground dwarf.

\subsection{Wil 1 stars at outlying velocities?}

To investigate the possibility that we missed one or more member stars
associated with Wil 1 (bound or unbound), we looked at the five
fRGB-colored stars at outlying velocities.  The equivalent widths of
their Na I absorption lines are similar to those of likely Wil 1
members at similar apparent magnitudes. However, the surface gravities
of faint red giant branch stars associated with Wil 1 are not very different
than those of MW foreground stars in that region of the CMD. Even
if this surface gravity indicator was robust for stars at these $g-r$
colors, it would not provide an effective discriminant. All five have
higher CaT equivalent widths than Wil 1 members at similar
apparent magnitudes.  We did not attempt this with MS/HB-colored stars
because the error bars on all measured parameters are too large to
provide a meaningful result.  Although there could possibly be one,
there is no obvious candidate for a star associated with Wil 1 in
the outlying tails of its velocity distribution.

\section{The Nature of Willman 1: A Disrupting Dwarf}\label{sec_dwarf}

\subsection{An [Fe/H] spread}\label{ssec_spread}

Despite the observed star-to-star variations in some of the light
elements, the Milky Way's globular clusters generally have not been
observed to have a significant dispersion in [Fe/H]
\citep[e.g.][]{carretta09a}.  Although
spectroscopic evidence exists for modest [Fe/H] spreads ($\sim$ 0.1 -
0.2 dex) in a small number of Milky Way globular clusters
\citep[e.g. M22 and M54][]{dacosta09a,carretta10a}, only $\omega$Cen
displays a large ($\sim$ 1 dex) star-to-star variation in [Fe/H]
\citep[e.g.][]{norris95a}. Owing to its unusual abundance
distribution, $\omega$Cen is regarded to be the remaining core of an
otherwise destroyed dwarf galaxy \citep{lee99a,bekki03a}.  Unlike the
globular clusters, all Milky Way dwarf galaxies are observed to have a
significant dispersion in their stars' [Fe/H].  Observing such a spread
is thus good evidence that an object self-enriched within a dark
matter halo---that it is a galaxy.

Assessing the evidence for a metallicity spread in Wil 1 is
challenging, because even minimal foreground contamination could lead
to the erroneous conclusion that a large [Fe/H] spread exists. In
\S 3.3, we applied a CaT W$'$ cut to reject possible
foreground stars, leaving 10 highly probable member stars of the
initial sample of 15 possible RGB members.  This selection left only
Stars 1 and 2 of the five brightest possible member stars.  We then
performed a detailed analysis in \S 3.4 to ensure that Stars 1 and 2
are true Wil 1 members.

Using the spectral synthesis method of KGS08, described in \S 2.5,
[Fe/H]$_{Star1}$ = $-1.73 \pm 0.12$ and [Fe/H]$_{Star2}$ = $-2.65 \pm
0.12$, indicating that Star 2 is 0.9 dex more iron-poor than Star 1.
To show these results aren't sensitive to S/N, we redo this analysis
after artificially reducing the S/N in each spectrum by a factor of
two with Gaussian random noise proportional to $\sqrt{\rm pixel
  variance}$. We find [Fe/H]$_{Star1}$ = $-1.77 \pm 0.12$ and
[Fe/H]$_{Star2}$ = $-2.86 \pm 0.20$.  The KGS08 technique also yields
[Ca/Fe]$_{Star1}$ = $-0.4 \pm 0.18$, and [Ca/Fe]$_{Star2}$ = $+0.13
\pm 0.28$.  This large spread of [Ca/Fe] abundances, with the more
metal-rich star having the lower [Ca/Fe], is consistent with a
scenario where Type Ia supernovae are controlling the chemical
enrichment of Wil 1 after its first generation of stars.  With only
two bright RGB stars, it is unlikely that these spreads sample the
full spread of abundances in Wil 1.

Stars 1 and 2 also have accurate photometry in SDSS DR7. We use the
photometric metallicity calibration of \citet{lenz98}: $l = -0.436 u +
1.129 g - 0.119 r - 0.574 i + 0.198$, valid in the range $0.5 < g-r <
0.8$. Stars with larger $l$ are more metal-poor. Stars 1 and 2 have
$l=0.137 \pm 0.032$ and $0.216 \pm 0.035$ respectively.  Although there
is no robust conversion between this Lenz photometric metallicity
statistic and stellar metallicity, we use their calculations to
estimate approximate photometric metallicities from the $l$
parameter. We fit a spline between $l$ and metallicity for the full
set of their calculations shown in their Figure 8. This fit yields
[M/H] estimates of $-1.8 \pm 0.3$ and $-3.0^{+0.7}_{-0.9}$ for Stars 1
and 2, respectively. Despite the significant random and (likely)
systematic errors in this process, these values are consistent
with our spectral synthesis measurements. 

\begin{figure}
%\epsscale{0.4}
\plotone{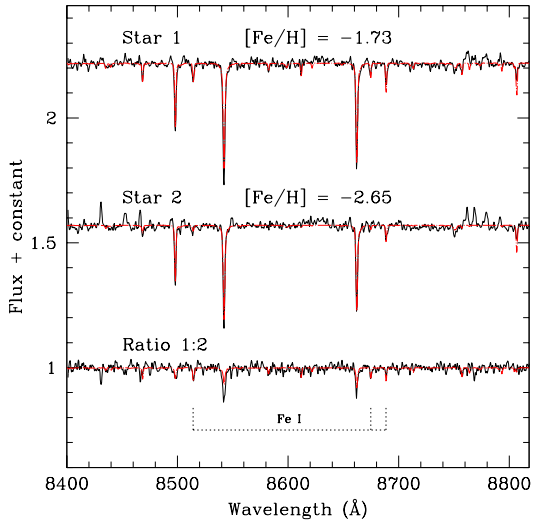}

\caption{Portions of the spectra of Stars 1 and 2 (as listed in Table
  2) around the \ion{Ca}{2} triplet features.  The bottom spectrum
  shows the ratio of these two stars; the clear features underscore
  the true difference in the abundances of these two Wil 1 member
  stars.  The spectral syntheses used to measure [Fe/H] are
  overplotted.}

\label{fig_spec}
\end{figure}

Figure~\ref{fig_spec} shows the spectra for these two stars in the
region around the CaT along with the ratio of the spectra.  The
spectral syntheses used to measure [Fe/H] are overplotted. Visually,
the individual spectra appear different, and weak lines can be seen in
Star 1's spectrum that are not visible in Star 2's.  The ratio of the
spectra clearly shows that metal lines are stronger in Star 1. Given
that the two stars have nearly identical luminosities and colors, this
visual comparison also demonstrates that Star 1 is significantly more
metal-rich than Star 2.

The scatter in W$'$ of the eight relatively faint Wil 1 RGB members
seen in Figure~\ref{fig_CaT} provides additional evidence for an
internal abundance spread in this object.  Taking the measurement
uncertainties on W$'$ into account, these stars have a mean W$'$ = 2.4
\AA\ and a dispersion = 0.7 \AA.

The method of KGS08 has been calibrated for stars with surface
gravities lower than 3.6.  Of the other eight probable Wil 1 RGB
members, only one has log $g < 3.6$ (Star 8).  We find that this star
has [Fe/H] = $-1.92 \pm 0.21$.  Averaging the [Fe/H] of Stars 1, 2,
and 8, we find a mean [Fe/H] of $-$2.1. At face value, this is
inconsistent with a simple linear metallicity-luminosity relation for
Milky Way dwarf satellites. For example, the relation of
\citet{kirby11a} predicts a mean [Fe/H] = $-2.7$. Conversely,
\citet{kirby11a}'s linear relation predicts that the typical
luminosity of a Milky Way satellite with [Fe/H] = $-$2.1 is $\sim 6.5
\times 10^4 L_{\odot}$.  This offset between the observed and expected
[Fe/H] of Wil 1 could result if the metallicity-luminosity relation
flattens as the lowest luminosities (as also suggested by observations
of Segue 1 \citealt{simon11a}).  Alternatively, it could result if Wil
1 has been stripped of a lot of its stellar mass, or if the mean
[Fe/H] based on three stars may not be an accurate reflection of the
average composition of stars in this system.

For the purpose of comparison, the [Fe/H] derived for Stars 1 and 2
based on the \citet{rutledge97} calibration of CaT W$'$ are $-1.97 \pm
0.17$ and $-2.07 \pm 0.17$, respectively. The [Fe/H] derived for Stars
1 and 2 based on the \citet{starkenburg10} calibration of CaT W$'$ are
$-2.36 \pm 0.20$ and $-2.64 \pm 0.29$, respectively. As expected, the
\citet{starkenburg10} values are lower metallicity, because the
\citet{rutledge97} [Fe/H] have been shown to systematically
overestimate [Fe/H] lower than -2.0 (see KGS08,
\citealt{starkenburg10} and references therein).  What is initially
unexpected is that the [Fe/H] values of Stars 1 and 2 as derived using
the CaT W$'$ as an [Fe/H] indicator underestimate the spread in [Fe/H]
between Stars 1 and 2.  This underestimate results from the large
spread in [Ca/Fe] between Stars 1 and 2 and underscores another
weakness of the CaT W$'$ approach to [Fe/H]: The CaT technique doesn't
account for differences in [Ca/Fe].  If we compare the [Ca/H] we
obtain using the relationship provided by \citet{starkenburg10} with
the [Ca/H] derived through spectral synthesis, we find excellent
agreement: [Ca/H]$_{Star1,Stark}$ = $-2.11 \pm 0.20$ and
[Ca/H]$_{Star1,Kirby}$ = $-2.13 \pm 0.13$, [Ca/H]$_{Star2,Stark}$ =
$-2.39 \pm 0.29$ and [Ca/H]$_{Star2,Kirby}$ = $-2.52 \pm 0.25$.

In summary, Wil 1 stars exhibit a substantial [Fe/H] spread.  The
[Fe/H] spread presented here is an underestimate if we have thrown out
true member stars as a result of our W$'$ member selection criterion.
We thus conclude that Wil 1 is a dwarf galaxy, or the remnants
thereof.

\subsection{Tentative evidence for a disturbed
  morphology}\label{ssec_distrib}

\begin{figure}
%\epsscale{0.5}
\plotone{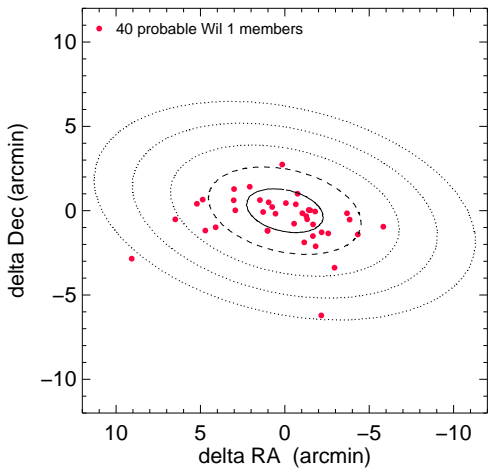}
 
\caption{The spatial distribution of the 40 Wil 1 member stars. For
  reference, dotted lines show 1, 2, 3, 4, 5 and 6$R_{\rm half}$
  around the center of the object, oriented at a position angle of
  77$^{\circ}$ \citep{martin08a}.
\label{fig_spatial2}}
\end{figure}

The spatial distribution of Wil 1 stars displays tentative evidence
for multi-directional features \citep[][although see
\citealt{martin08b} and \citealt{walsh08a} for discussion of shot
noise and morphology]{willman06a,martin07a} and has a moderately high
ellipticity \citep[$e = 0.47 \pm 0.08$]{martin08b}. Additionally, the
spatial distribution of Wil 1's spectroscopic member stars provides
tenuous evidence for an extended spatial
distribution. Figure~\ref{fig_spatial2} shows that two of the 40 Wil 1
spectroscopic members lie at $\sim 5 R_{\rm half}$.  Only 1\% of the
stars following an exponential distribution lie beyond 4 $R_{\rm
half}$.  To test the likelihood of two outlying members occurring by
chance in a system well described by an exponential distribution, we
ran Monte Carlo simulations of the expected distribution of Wil 1
stars, assuming the \citep{martin08b} structural parameters ($R_h =
2.3\arcmin$; $\epsilon = 0.47$) and the target efficiency in Figure
5. In only 1.1\% of simulations were there two or more stars between 4
and 6 $R_h$. Using the simulations described in \S 3.2, we find it
unlikely that either of these stars is a contaminant.  For example,
there is a $< 2$\% chance that they are both foreground stars.  These
probabilities all depend on the assumption that Wil 1 is perfectly
described by an exponential function (a larger fraction of a Plummer
galaxy's light resides at large distance) and that its scale length is
exactly that measured by
\citep{martin08b}.  We thus consider these distant members to provide
only tenuous evidence that Wil 1 may have an excess of stars at large
radii, which could imply ongoing tidal stripping.

Each individual piece of evidence for a possible disturbed morphology
of Wil 1 is not remarkable.  When combined, they provide a reasonable
(but still tentative) basis for believing that Wil 1 has been
structurally disturbed, perhaps by the tidal field of the Milky Way.

%%%%%%%%%%%%%%%%%%%%%%%%%%%%%%%%%%
% Figure:  MEMBER DISTRIBUTION
%%%%%%%%%%%%%%%%%%%%%%%%%%%%%%%%%%

\section{The Kinematics of Wil 1}\label{sec_kin}

\subsection{Dynamical evidence for a high dark matter content?}

We begin by characterizing the global systemic velocity and velocity
dispersion of Wil 1 with a single number, using the maximum likelihood
method of \citet{walker06a}. This method assumes an intrinsic Gaussian
distribution for the line-of-sight velocities of Willman 1, and adds
in quadrature to the intrinsic dispersion the measurement error on
each star. Though it is possible that the system could be bound, in
dynamical equilibrium and have a distribution function described by
mild deviations from Gaussianity at all radii, the fact that the
measurement errors are independent and Gaussian makes the Gaussian
form of the likelihood an adequate general description of the
system. Using the sample of 40 probable member stars, $v_{sys} =
-$14.1 \kms $\pm$ 1.0 \kms and $\sigma_v = $ 4.0 \kms $\pm$ 0.8 \kms.
If we also include the 5 more metal-rich stars flagged as probable
contaminants in \S3.3, $v_{sys} = -$12.8 \kms $\pm$ 1.0 \kms and
$\sigma_v = $ 4.8 \kms $\pm$ 0.8 \kms.  3$\sigma$ clipping removes no
stars from either the 40 or the 45 star sample.

Given the line-of-sight velocity dispersion as calculated above, and
assuming dynamical equilibrium, it is straightforward to estimate the
mass of Wil 1. \citet{wolf10a} use the sample of 40 probable
member stars presented here and determine the mass within the
half-light radius of of $\sim 3.9^{+2.5}_{-1.6} \times 10^5$
$M_{\odot}$. This mass is relatively insensitive to the modeling of
the velocity anisotropy profile and the parameterization of the light
distribution of Wil 1.  In terms of the central mass-to-light ratio,
(M/L)$_{V}$, of 770$^{+930}_{-440}$.  A calculation including the 5
apparently metal-richer stars would imply a 40\% larger mass.  Even if
all of the other assumptions were robust, we believe this higher
inferred mass would be erroneous for the reasons given in \S3.3.

\subsection{An unusual and inconclusive kinematic
  distribution}\label{ssec_unusual}

\begin{figure*}[t!]
\epsscale{0.8}
\plotone{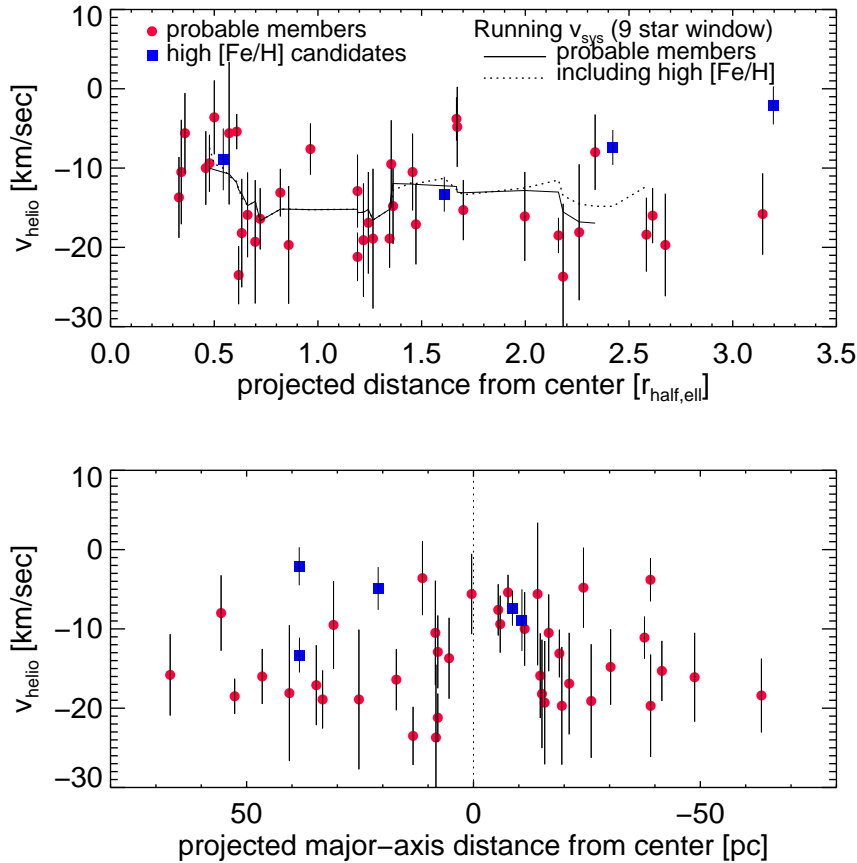}

\caption{Heliocentric line-of-sight velocities of probable Wil 1
  member stars high CaT W$'$ stars classified as possible interlopers
  in \S 3.3.  Top panel: Velocities as a function of 2D elliptical
  distance from Wil 1's center, including the 38 probable members (of
  40) and the 4 possible high [Fe/H] interlopers (of 5) with $r_{\rm
  ell} < 3.5 r_{\rm half}$.  Overplotted is the systemic velocity of
  Wil 1, calculated in a 9 star running window, calculated both with
  and without the possible interloper stars. Lower panel: The
  line-of-sight velocities of stars as a function of their 1D
  projected distance along Wil 1's major axis, including the 39
  probable members and the 5 possible high [Fe/H] interlopers with
  $d_{\rm major} <$ 3.5 80 pc.  The dotted line highlights the center
  of Wil 1.  These panels show that neither is Wil 1's systemic
  velocity well described by a single value, nor is there clear
  evidence of rotation.}
 
\label{fig_vdistr}

\end{figure*}

We now investigate the kinematic distribution of Wil 1 stars in more
detail.  Is it even reasonable to characterize the systemic velocity
and velocity dispersion of Wil 1 with a single number?
Figure~\ref{fig_vdistr} shows the line-of-sight velocities of the
probable member stars (red circles) as a function of distance.  The
top panel shows elliptical distance from the center and the bottom
panel shows distance along the major-axis relative to the center.
Blue squares in both panels show the 4 stars within 3.5$r_{\rm half}$ with
CaT W$'$ $>$ 3.9 \AA\ that we classified as likely interloper stars. Two
member stars and one CaT W$'$ $>$ 3.9 \AA\ star lie beyond the edge of
both panels.  We exclude these distant stars from further kinematic
analysis because they may be remnants that have been stripped from 
the main body of the object (see \S4.2).

Stars in the top panel of Figure~\ref{fig_vdistr} display an unusual
kinematic distribution: central stars have velocities systematically
offset from those of more distant member stars.  The solid line shows
the systemic velocity of Wil 1 calculated in a running window of nine
stars, not including those identified with high [Fe/H].  The running
$v_{\rm sys}$ rapidly decreases by $\sim$ 8 \kms\ from the center to
the outskirts of Wil 1. The dotted line of Figure~\ref{fig_vdistr}
shows the same as the solid line, but including the likely interloper
stars.  This line shows that our result is not affected by our
criterion for interloper identification. It is tempting to invoke a
tidally-disrupting scenario may provide an explanation of this
behavior in the systemic velocity, however this velocity distribution
might not be easily explained by existing models of tidally stripped
dwarf galaxies.  For example, \citet{klimentowski09} show that
different viewing angles of a stripped dwarf galaxy are expected to
reveal a symmetric velocity distribution of unbound stream stars.  We
note that the details of this systemic velocity profile depends on the
center used for Wil 1 and on the running window size.  While we
calculated Wil 1's center (\S 2.1) as accurately as we could and the
window size used here was not fine tuned to produce this result, if
our center is inaccurate or if we choose a larger running window, we
may not see such a striking shift in systemic velocity with
distance. Given the relatively small kinematic sample from Wil 1 and
its extreme faintness it is difficult to conclusively interpret this
systemic velocity variation.

%The relatively discrepant velocities of 3 of the 4
%blue squares relative to likely members as similar projected distances
%from the Wil 1 center (seen in the top panel) again supports our
%hypothesis that they are true interlopers.

The bottom panel of Figure~\ref{fig_vdistr} shows that this unusual
kinematic distribution is not obviously a result of ordered rotation.
Visually, a model with velocity that is a linear function of major
axis position (as expected for ordered rotation) does not provide a
good fit to these data. Additionally, even if the system were rotating
with a reasonable velocity, this signal would not be able to be
discerned with such a small sample of discrete velocities
\citep{strigari10a}.

The variation in the systemic velocity with radius makes it difficult
to robustly define the velocity dispersion of Wil 1. In fact we find
the velocity dispersion profile of Wil 1 to be highly dependent on the
running window or binsize used to calculate the profile.  We therefore
choose not to show a dispersion profile here. We note for the 40 Wil 1
members that i) the velocity dispersion of the innermost 8 stars stars
is equal to 0 \kms\, with an uncertainty of 2.1 \kms\ ; ii) the
velocity dispersion of the 9 stars with distances between 1.0 and 1.5
$r_{\rm ell}$ is also equal to 0 \kms\ , with an uncertainty of 2.5
\kms\ ; and iii) the velocity dispersion of the outermost 9 or ten
stars shown in Figure 8 is equal to 0 \kms\ , with an uncertainty of
2.1 \kms\ .  The details of this result is sensitive to the center
used for Wil 1.

We now explicitly address each of the assumptions necessary for using
the kinematics to interpret a high mass-to-light ratio for Wil
1. While these assumptions may be reasonable for many of the Milky
Way's companions, it is not yet clear that they are reasonable
assumptions for Wil 1.

\begin{itemize}

\item {\it All 40 stars are physically associated with Wil 1} 

  As stated in \S3, There may be a small number of Milky Way stars
  remaining in our sample of 40 likely Wil 1 members, likely among its
  apparent MS members.  The velocity dispersions of the Milky Way
  halo and thick disk are larger than that measured for Wil 1, so
  stars belonging to them may somewhat artificially inflate the global
  velocity dispersion measured for Wil 1.

\item {\it All stars associated with Wil 1 are bound and in dynamical
    equilibrium}

  Our sample of 40 contains too few stars to robustly check whether
  the velocity distribution of stars in Wil 1 is consistent with
  dynamical equilibrium.  Its unusual spatial and velocity
  distributions (Figure~\ref{fig_vdistr} and discussed above), excess
  of spectroscopic members at large distance (\S 3.4), and its
  relatively high [Fe/H] for its luminosity (\S5) may indicate that
  Wil 1 is a disrupted or disrupting object.  If this were the case
  then the assumption of dynamical equilibrium would not be valid. It
  is plausible that the 40 member stars reported here may contain some
  (many) unbound stars.

\item {\it Contribution to velocities from binary stars}

Given the small velocity dispersion that we have measured, and the
fact that the measurement uncertainities are similar to the measured
velocity dispersion, it is possible that binary stars may be inflating
the global line-of-sight velocity dispersion measured for Wil 1 stars.
\citet{minor10} simulated the effect of binary stars observed velocity
dispersions of dwarf spheroidal galaxies.  They concluded that, while
binaries do inflate the observed velocity dispersion of systems such
as Wil 1, they are not expected to have inflated the velocity
dispersion of a system with $\sigma_{obs}$ = 4 km/sec by more than 0.8
km/sec over its $\sigma_{intrinsic}$.  Alternative binary models have
suggested that the observed dispersions of systems with intrinsic
dispersions of only a few tenths of a km/sec could be even more
affected by the presence of binaries (A. McConnachie, private
communication). Both modeling and repeated observations of individual
Wil 1 member stars would be necessary to definitively conclude the
effect of binaries on its observed velocity dispersion.

\item {\it Symmetric velocity distribution} 

  The distribution of line-of-sight velocities, relative to the mean
  velocity, is symmetric for any equilibrium galaxy model. While
  noting again that our sample size is too small to reach any
  statistically-robust conclusion, there are initial hints of asymmetry
  in the line-of-sight velocity distribution (See Fig. 8).  If this
  asymmetric distribution persists with future data sets then it would
  support the hypothesis that Wil 1 is not a dynamically equilibrated
  system.

\end{itemize}

\section{Discussion and Conclusions}

The DEIMOS spectroscopic study presented here has revealed several new
insights into the unusual Wil 1 object and has underscored the
importance of careful foreground characterization when studying the
least luminous Milky Way companions. We have shown that the velocity,
color, and magnitude overlap of Wil 1's stellar population with
foreground Milky Way stars make this object particularly difficult to
study. We thus performed detailed Monte Carlo simulations to calculate
the possible foreground contamination. We used CaT W$'$ plus color
information to identify a total of 5 likely interloper stars out of
the 45 possible spectroscopic members selected in \S3.1.  The high
interloper fraction we estimated for Wil 1's brightest candidate RGB
member stars does not imply that Wil 1's true luminosity is less than
that derived in past studies.  Photometric studies of ultra-faint
dwarfs \citep[e.g.][]{martin07a,sand09a,munoz10a} typically use a
statistical definition of their luminosity rather than adding up the
light emitted from individual stars.

In concert with the Besancon Galaxy model, spectra from the SEGUE
database helped confirmed the similarity of two of the five flagged
interloper stars to Milky Way stars with similar color, magnitude and
velocity. SEGUE spectra also provided strong support for the presence
of two true bright red giant branch Wil 1 members (Stars 1 and 2 in
Table 2, discussed in \S3.3.1).  SEGUE may thus provide a valuable
resource for future studies that aim to eliminate interloper stars
from spectroscopic samples.

The mean [Fe/H] of Wil 1's three confirmed RGB members with log $g <
3.6$ is $-$2.1, with a difference of 0.9 dex between the most
metal-poor and metal-rich star.  We found [Ca/Fe]$_{Star1}$ = $-0.4
\pm 0.18$, and [Ca/Fe]$_{Star2}$ = $+0.13 \pm 0.28$ for the brightest
two stars, with the more Fe-rich star having the lower [Ca/Fe].  As
discussed in \S~\,\ref{ssec_spread}, we interpret the large [Fe/H]
difference between these stars to demonstrate that Wil 1 is (or once
was) a dwarf galaxy, rather than a star cluster.  With $r_0 \sim 18.1$
for the brighter two of these RGB members, they are good targets for
high-resolution spectroscopic follow-up to investigate their detailed
abundance patterns.  Because Star 2 was observed by \citet{siegel08a},
Star 1 is the top priority for follow-up.

The kinematic distribution of Wil 1 is unlike the distribution yet seen
in any of the Milky Way's satellites.  Its inner 9 spectroscopic
member stars have radial velocities offset by 8 \kms\ from its 29 more
distant members (excludes the two members more distant than 3.5
$r_{\rm half}$).  Neither published models of tidally disturbed
satellites nor ordered rotation provide an easy explanation for this
distribution.  We emphasize that the exact character of Wil 1's
systemic velocity and velocity dispersion profile depends sensitively
on the running window and the exact center used.  We thus use this
present dataset to highlight the unusual nature of Wil 1's kinematics
rather than to present definitive conclusions.

Wil 1's possible disturbed morphology and tentative excess of
spectroscopic members at large distance relative to that of an
undisturbed exponential distribution (\S~\ref{ssec_distrib}) suggests
that it may have been stripped of a substantial fraction of its
stellar component. Some models of the tidal evolution of dark matter
dominated satellites suggest that Wil 1 should presently have a high
M/L even if it has suffered substantial tidal evolution
\citep{penarrubia08a}.  However, because its dark mass content cannot
be well constrained given the reasons articulated above, appropriate
caution must be taken when attempting to compare this object to the
Milky Way's other satellites or attempting use it to constrain the
particle nature of dark matter.

If Wil 1 has been severely stripped of stars and the line-of-sight
velocities of its stars do not trace the underlying gravitational
potential, then why does it happen to lie on the $M_{\rm
dynamical}/L_V$ vs $L_V$ relationship observed for Milky Way dwarf
satellites \citep{geha09a,wolf10a}?  Perhaps this is a coincidence, or
perhaps the velocities of Wil 1's stars do actually trace its
gravitational potential, despite its overall unusual kinematic
distribution.  Only numerical models aimed specifically to reproduce
Wil 1's properties may illuminate which answer is correct. Searches
for dwarfs that can reveal the presence (or lack) of Willman
1-luminosity objects at a wide range of halo distances will be needed
to know for certain the role of environment in shaping the
luminosities of the tiniest Milky Way satellites.

%\citet{willman06} hypothesized that Wil 1 may be near the apocenter of
%its orbit to explain its irregular morphology.  To convert $v_{sys} =
%-$14.1 \kms into Galactocentric velocity, we use the values of (U,V,W)
%from Xue et al (2008): (10.0,5.2,7.2) and v$_{lsr}$ = 220 km/s and
%find v$_{GSR}$ = v$_{hel}$ + 46.0 \kms = 31.9 /kms. If we use the
%recently favored higher value of v$_{lsr} \sim 245$ \kms
%\citep{bovy09a}, then this relation changes to v$_{GSR}$ = v$_{hel}$ +
%50.8 \kms = 36.7 /kms.  Wil 1 may thus be close to apocenter.

\acknowledgments

BW acknowledges support from NSF AST-0908193 and thanks Ewa Lokas,
Anil Seth, Joe Wolf, and Gail Gutowkski for interesting and helpful
conversations during the preparation of this paper. We also thank the
anonymous referee for providing thoughtful suggestions that resulted
in substantial improvement of this manuscript. MG acknowledges support
from NSF AST-9008752.  ENK, J. Strader, and LES acknowledge support
provided by NASA through Hubble Fellowship grants HST-HF-01233.01,
HST-HF-51237.01, and HST-HF-01225.01 respectively, awarded by the
Space Telescope Science Institute, which is operated by the
Association of Universities for Research in Astronomy, Inc., for NASA,
under contract NAS 5-26555.  We thank David W. Hogg and Morad Masjedi
for obtaining the Wil 1 observations at KPNO in 2005. Some of the data
presented herein were obtained at the W.M. Keck Observatory, which is
operated as a scientific partnership among the California Institute of
Technology, the University of California and the National Aeronautics
and Space Administration. The Observatory was made possible by the
generous financial support of the W.M. Keck Foundation. This research
has also made use of NASA's Astrophysics Data System Bibliographic
Services.

%\bibliographystyle{../apj}
%\bibliography{../../../master}

\clearpage

\begin{deluxetable}{lccccccc}
\tabletypesize{\scriptsize}
\tablecaption{Keck/DEIMOS Multi-Slitmask Observations of Willman1}
\tablewidth{0pt}
\tablehead{
\colhead{Mask} &
\colhead{$\alpha$ (J2000)} &
\colhead{$\delta$ (J2000)} &
\colhead{PA} &
\colhead{$t_{\rm exp}$} &
\colhead{\# of slits} &
\colhead{\% useful} \\
\colhead{Name}&
\colhead{(h$\,$:$\,$m$\,$:$\,$s)} &
\colhead{($^\circ\,$:$\,'\,$:$\,''$)} &
\colhead{(deg)} &
\colhead{(sec)} &
\colhead{}&
\colhead{spectra}
}
\startdata
Wil1\_1  & 10:49:23 & +51:01:20  &  75 & $5 \times 1800$ & 110 & 58\%\\
Wil1\_2  & 10:49:40 & +51:01:57  & 110 & $5 \times 1800$ &  94 & 45\%\\
Wil1\_3  & 10:49:11 & +51:02:16  &  20 & $3 \times 1800$ &  92 & 50\%\\
Wil1\_4  & 10:49:24 & +51:01:02  &  -5 & $3 \times 1800$ & 127 & 7\%
\enddata
\tablecomments{Right ascension, declination, position angle and total exposure
time for each Keck/DEIMOS slitmask in Willman~1.  The final two
columns refer to the total number of slitlets on each mask and the
percentage of those slitlets for which a redshift was measured.  Mask
4 has a low efficiency because many faint stars were targeted.}
\end{deluxetable}

\begin{deluxetable}{lccccccc}
\tabletypesize{\scriptsize}
\tablecaption{Data for the 45 candidate Willman 1 members}
\tablewidth{0pt}
\tablehead{
\colhead{ID} &
\colhead{$\alpha$ (J2000)} &
\colhead{$\delta$ (J2000)} &
\colhead{$r$} &
\colhead{$(g-r)$} &
\colhead{$v_{\rm helio}$} &
\colhead{S/N} &
\colhead{CaT EW} \\
\colhead{}&
\colhead{(h$\,$ $\,$ m$\,$ $\,$s)} &
\colhead{($^\circ\,$ $\,'\,$ $\,''$)} &
\colhead{(mag)} &
\colhead{(mag)} &
\colhead{(\kms)} &
\colhead{ } &
\colhead{\AA\ } }
\startdata
   1 &  10 49 18.06 & +51 02 16 & 18.13 &  0.57 &   $-$5.4 $\pm$  2.2 &   84.1&  1.63 $\pm$  0.41 \\
   2 &  10 49 52.51 & +51 03 42 & 18.14 &  0.58 &  $-$18.5 $\pm$  2.2 &   85.5&  1.41 $\pm$  0.41 \\
   3\tablenotemark{a} &  10 49 42.87 & +51 04 22 & 18.19 &  0.65 &  $-$13.3 $\pm$  2.2 &   88.1&  4.24 $\pm$  0.39 \\
   4\tablenotemark{a} &  10 49 12.40 & +51 05 43 & 18.21 &  0.63 &   $-$7.4 $\pm$  2.2 &   72.3&  4.65 $\pm$  0.41 \\
   5 &  10 49 07.79 & +50 56 50 & 18.35 & $-$0.04 &  $-$11.1 $\pm$  2.7 &   34.3& -- \\
   6\tablenotemark{a} &  10 49 48.53 & +51 00 32 & 19.05 &  0.39 & $-$2.1 $\pm$  2.4 &   21.6&  3.65 $\pm$  0.46 \\
   7 &  10 49 13.13 & +51 02 32 & 19.81 & $-$0.32 &  $-$18.2 $\pm$  6.8 &   18.5& -- \\
   8 &  10 49 17.41 & +51 03 25 & 20.07 &  0.44 &   $-$9.4 $\pm$  3.6 &    9.4& 1.39 $\pm$  0.46 \\
   9 &  10 49 10.11 & +51 03 00 & 20.50 &  0.42 &  $-$13.1 $\pm$  3.0 &   16.3&  2.38 $\pm$  0.40 \\
  10\tablenotemark{a} &  10 49 15.95 & +51 02 26 & 20.54 &  0.45 &   $-$8.9 $\pm$  3.9 &    9.8&  4.46 $\pm$  0.41 \\
  11\tablenotemark{a} &  10 49 24.97 & +51 09 23 & 20.60 &  0.44 &   $-$4.9 $\pm$  2.7 &   16.8&  4.51 $\pm$  0.39 \\
  12 &  10 49 21.14 & +51 03 29 & 20.82 &  0.36 &   $-$5.6 $\pm$  5.1 &    5.6& 0.75 $\pm$  0.44 \\
  13 &  10 48 58.10 & +51 02 53 & 20.86 &  0.46 &   $-$3.8 $\pm$  2.7 &    8.3&  2.79 $\pm$  0.40 \\
  14 &  10 49 40.82 & +51 03 39 & 20.94 &  0.37 &  $-$18.9 $\pm$  3.7 &   15.3&  2.60 $\pm$  0.40\\
  15 &  10 49 30.94 & +51 03 40 & 20.96 &  0.34 &  $-$16.4 $\pm$  3.9 &   13.2&  2.21 $\pm$  0.41 \\
  16 &  10 49 16.75 & +51 04 03 & 21.00 &  0.34 &   $-$7.6 $\pm$  3.2 &   10.6&  3.69 $\pm$  0.44 \\
  17 &  10 50 02.85 & +51 02 32 & 21.27 &  0.43 &  $-$15.8 $\pm$  5.1 &    5.4&  3.12 $\pm$  0.41 \\
  18 &  10 50 19.29 & +51 00 12 & 21.39 &  0.34 &  $-$24.2 $\pm$  2.7 &    7.5& -- \\
  19 &  10 48 57.12 & +51 02 31 & 21.42 &  0.26 &  $-$15.3 $\pm$  3.9 &    7.3& -- \\
  20 &  10 49 28.06 & +51 01 51 & 21.45 &  0.22 &  $-$21.2 $\pm$  3.1 &    3.5& -- \\
  21 &  10 49 28.06 & +51 01 51 & 21.45 &  0.22 &  $-$12.9 $\pm$  4.6 &    3.4& -- \\
  22 &  10 49 12.55 & +51 03 05 & 21.49 &  0.25 &  $-$15.9 $\pm$  5.3 &    7.0& -- \\
  23 &  10 49 26.31 & +51 03 16 & 21.49 &  0.22 &  $-$10.5 $\pm$  6.7 &    8.5& -- \\
  24 &  10 49 25.06 & +51 02 52 & 21.54 &  0.25 &  $-$13.7 $\pm$  5.1 &    4.6& -- \\
  25 &  10 49 07.67 & +51 01 46 & 21.56 &  0.23 &  $-$19.1 $\pm$  7.2 &    6.9& -- \\
  26 &  10 48 44.41 & +51 02 06 & 21.57 &  0.28 &  $-$18.4 $\pm$  4.7 &    7.3& -- \\
  27 &  10 49 51.52 & +51 01 52 & 21.58 &  0.36 &  $-$16.0 $\pm$  3.5 &    8.3& -- \\
  28 &  10 49 34.77 & +51 04 28 & 21.60 &  0.27 &  $-$18.9 $\pm$  8.8 &    7.2& -- \\
  29 &  10 49 09.93 & +51 00 56 & 21.63 &  0.26 &   $-$4.8 $\pm$  5.0 &    6.4& -- \\
  30 &  10 49 29.73 & +51 02 58 & 21.64 &  0.21 &  $-$23.5 $\pm$  3.7 &    6.1& -- \\
  31 &  10 49 54.69 & +51 03 27 & 21.66 &  0.27 &   $-$8.0 $\pm$  4.8 &    4.0& -- \\
  32 &  10 49 14.23 & +51 01 10 & 21.66 &  0.23 &  $-$10.5 $\pm$  4.8 &    7.6& -- \\
  33 &  10 49 27.64 & +51 03 32 & 21.75 &  0.22 &   $-$3.6 $\pm$  4.8 &    3.8& -- \\
  34 &  10 49 05.17 & +51 01 42 & 21.80 &  0.22 &  $-$14.8 $\pm$  4.9 &    6.6& -- \\
  35 &  10 49 22.53 & +51 05 47 & 22.04 &  0.27 &  $-$23.7 $\pm$  9.2 &    2.4& -- \\
  36 &  10 49 11.98 & +51 03 04 & 22.13 &  0.28 &  $-$19.3 $\pm$  7.8 &    3.1& -- \\
  37 &  10 49 02.81 & +50 59 40 & 22.20 &  0.27 &  $-$19.7 $\pm$  6.5 &    4.8& -- \\
  38 &  10 49 40.69 & +51 04 19 & 22.26 &  0.24 &  $-$17.1 $\pm$  5.0 &    4.4& -- \\
  39 &  10 49 14.93 & +51 02 53 & 22.44 &  0.28 &  $-$10.0 $\pm$  4.6 &    3.6& -- \\
  40 &  10 49 13.44 & +51 02 43 & 22.48 &  0.36 &   $-$5.6 $\pm$  9.0 &    3.1& -- \\
  41 &  10 49 10.91 & +51 02 14 & 22.61 &  0.29 &  $-$19.7 $\pm$  7.4 &    2.3& -- \\
  42 &  10 49 47.62 & +51 02 03 & 22.62 &  0.28 &  $-$18.1 $\pm$  8.6 &    3.5& -- \\
  43 &  10 48 54.03 & +51 01 38 & 22.63 &  0.31 &  $-$16.1 $\pm$  5.6 &    3.1& -- \\
  44 &  10 49 40.22 & +51 03 04 & 22.89 &  0.29 &   $-$9.5 $\pm$  5.5 &    2.9& -- \\
  45 &  10 49 10.96 & +51 01 32 & 22.93 &  0.32 &  $-$16.9 $\pm$  6.4 &    2.4& -- 
\enddata

\tablecomments{S/N is the median per pixel signal-to-noise for each
  star.  Velocity error bars were determined from measurement overlaps
  as discussed in \S\,\ref{ssec_rvel}. We supply the CaT EW only for
  stars possibly in the fRGB or bRGB populations of Wil 1.}
\tablenotetext{a}{Star flagged as a likely non-member by the CaT EW
  $<$ 2.3 \AA\ or color criterion described in \S3.3.}

\end{deluxetable}

\begin{deluxetable}{lcccccc}
\tabletypesize{\scriptsize}
\tablecaption{Data for the 52 Milky Way foreground stars in our sample}
\tablewidth{0pt}
\tablehead{
\colhead{ID} &
\colhead{$\alpha$ (J2000)} &
\colhead{$\delta$ (J2000)} &
\colhead{$r$} &
\colhead{$(g-r)$} &
\colhead{$v_{\rm helio}$} &
\colhead{S/N} \\
\colhead{}&
\colhead{(h$\,$ $\,$ m$\,$ $\,$s)} &
\colhead{($^\circ\,$ $\,'\,$ $\,''$)} &
\colhead{(mag)} &
\colhead{(mag)} &
\colhead{(\kms)} &
\colhead{ }}
\startdata
  46 &  10 50 15.74 & +51 02 22 & 15.31 &  0.65 &  $-$24.4 $\pm$  2.2 &  227.1 \\
  47 &  10 49 09.55 & +50 54 53 & 15.50 &  0.67 &   16.7 $\pm$  2.3 &  220.2 \\
  48 &  10 49 07.46 & +51 04 06 & 16.27 &  0.85 &    6.4 $\pm$  2.2 &  172.9 \\
  49 &  10 48 38.69 & +51 00 26 & 16.95 &  0.52 & $-$119.7 $\pm$  2.2 &  132.6 \\
  50 &  10 49 54.92 & +51 00 40 & 17.38 &  0.65 &  $-$53.8 $\pm$  2.2 &   87.8 \\
  51 &  10 49 27.20 & +50 59 26 & 18.57 &  0.55 &  155.1 $\pm$  2.3 &   43.3 \\
  52 &  10 48 45.47 & +50 58 40 & 19.17 &  0.57 & $-$171.9 $\pm$  2.3 &   37.0 \\
  53 &  10 48 47.90 & +50 57 17 & 19.33 &  0.40 &   14.8 $\pm$  2.3 &   34.4 \\
  54 &  10 49 13.10 & +51 06 26 & 19.50 &  0.86 &  $-$24.5 $\pm$  2.6 &   19.0 \\
  55 &  10 50 02.13 & +51 02 04 & 19.62 &  1.26 &  $-$29.5 $\pm$  2.2 &   47.5 \\
  56 &  10 49 21.51 & +51 08 26 & 20.03 &  1.42 &   13.2 $\pm$  2.2 &   78.3 \\
  57 &  10 50 19.02 & +50 59 18 & 20.03 &  0.41 & $-$249.3 $\pm$  2.4 &   20.3 \\
  58 &  10 50 13.69 & +51 00 06 & 20.04 &  1.12 & $-$205.0 $\pm$  2.3 &   23.3 \\
  59 &  10 49 12.90 & +51 06 17 & 20.17 &  0.47 &   14.3 $\pm$  2.4 &   18.4 \\
  60 &  10 49 28.48 & +51 01 33 & 20.27 &  1.41 &  $-$23.3 $\pm$  2.3 &   24.6 \\
  61 &  10 48 55.77 & +51 01 19 & 20.32 &  1.14 & $-$157.8 $\pm$  2.5 &   26.4 \\
  62 &  10 50 05.69 & +51 01 54 & 20.33 &  1.40 &  $-$10.7 $\pm$  2.2 &   28.7 \\
  63 &  10 49 56.07 & +51 00 00 & 20.61 &  1.40 &    6.2 $\pm$  2.2 &   24.0 \\
  64 &  10 49 22.03 & +50 59 05 & 20.73 &  0.32 &  $-$36.3 $\pm$  3.6 &    8.2 \\
  65 &  10 49 20.02 & +51 04 58 & 20.83 &  1.35 &   60.7 $\pm$  2.3 &   31.3 \\
  66 &  10 49 34.57 & +51 02 39 & 20.85 &  0.44 &  $-$52.8 $\pm$  4.8 &    7.9 \\
  67 &  10 49 13.34 & +51 04 28 & 20.85 &  1.39 &  $-$21.9 $\pm$  2.3 &   14.7 \\
  68 &  10 49 37.15 & +51 02 37 & 20.92 &  1.03 &  $-$69.5 $\pm$  2.4 &   15.0 \\
  69 &  10 50 09.54 & +50 59 22 & 21.06 &  1.33 & $-$205.3 $\pm$  2.4 &   14.6 \\
  70 &  10 50 11.50 & +51 01 11 & 21.21 &  0.50 &   39.9 $\pm$  3.8 &    6.5 \\
  71 &  10 49 39.43 & +51 02 03 & 21.28 &  1.19 &  $-$69.1 $\pm$  2.3 &   31.3 \\
  72 &  10 49 31.01 & +51 01 33 & 21.34 &  0.41 &  $-$78.9 $\pm$  4.0 &   10.3 \\
  73 &  10 49 53.75 & +51 00 43 & 21.35 &  1.37 &  $-$52.2 $\pm$  2.3 &   14.9 \\
  74 &  10 49 03.74 & +51 00 38 & 21.38 &  1.32 &  $-$49.0 $\pm$  2.5 &   18.1 \\
  75 &  10 49 13.88 & +51 05 12 & 21.47 &  1.40 &  $-$14.7 $\pm$  3.0 &    7.9 \\
  76 &  10 49 03.41 & +51 00 48 & 21.53 &  0.14 &    7.4 $\pm$  5.2 &    6.8 \\
  77 &  10 49 21.25 & +51 09 48 & 21.79 &  1.33 & $-$100.1 $\pm$  2.4 &   19.2 \\
  78 &  10 49 52.91 & +51 03 14 & 21.80 &  1.30 &  $-$37.9 $\pm$  3.2 &    6.7 \\
  79 &  10 48 48.74 & +51 03 16 & 21.82 &  0.46 &   65.9 $\pm$  4.7 &    6.3 \\
  80 &  10 49 07.25 & +51 02 12 & 21.87 &  1.44 &   50.6 $\pm$  2.4 &   22.9 \\
  81 &  10 49 30.19 & +51 07 23 & 21.90 &  1.34 &  $-$59.7 $\pm$  4.3 &   15.4 \\
  82 &  10 49 06.01 & +51 02 51 & 22.11 &  0.58 &   68.1 $\pm$  8.8 &    4.2 \\
  83 &  10 49 15.36 & +51 01 05 & 22.15 &  1.50 &  $-$41.4 $\pm$  2.5 &   24.1 \\
  84 &  10 49 14.03 & +51 08 44 & 22.22 &  1.30 &   $-$4.9 $\pm$  4.4 &    6.4 \\
  85 &  10 49 15.46 & +51 05 51 & 22.23 &  1.35 &  $-$73.4 $\pm$  2.6 &   18.3 \\
  86 &  10 49 20.72 & +51 01 41 & 22.23 &  0.25 &    8.7 $\pm$  7.2 &    3.5 \\
  87 &  10 49 39.60 & +51 02 26 & 22.25 &  1.03 &   87.3 $\pm$  5.3 &   16.6 \\
  88 &  10 49 24.43 & +51 09 14 & 22.38 &  1.40 &   50.4 $\pm$  2.6 &   15.3 \\
  89 &  10 50 14.71 & +50 59 46 & 22.42 &  0.82 &  $-$32.9 $\pm$  7.1 &    3.4 \\
  90 &  10 49 24.62 & +51 07 56 & 22.56 &  1.02 &   32.9 $\pm$  3.3 &    7.9 \\
  91 &  10 48 54.87 & +51 00 07 & 22.57 &  1.49 &  $-$54.4 $\pm$  3.4 &   21.2 \\
  92 &  10 49 30.07 & +51 08 19 & 22.61 &  1.25 &  $-$65.2 $\pm$  3.5 &    8.0 \\
  93 &  10 50 14.66 & +51 02 16 & 22.61 &  1.39 &   19.8 $\pm$  6.4 &    5.1 \\
  94 &  10 50 07.85 & +51 02 07 & 22.77 &  1.29 & $-$200.9 $\pm$  5.0 &    6.7 \\
  95 &  10 49 05.30 & +51 00 24 & 22.89 &  1.20 &  124.3 $\pm$  4.7 &   12.1 \\
  96 &  10 48 57.47 & +50 57 54 & 23.05 &  1.15 &  $-$10.0 $\pm$  3.1 &    9.2 \\
  97 &  10 49 38.33 & +51 01 38 & 23.08 &  0.44 &    5.7 $\pm$  7.8 &    2.1 
\enddata

 \tablecomments{S/N is the median per pixel
  signal-to-noise for each star.  Velocity error bars were
  determined from measurement overlaps as discussed in
  \S\,\ref{ssec_rvel}.}
\end{deluxetable}

\begin{deluxetable}{lccc}
\tabletypesize{\scriptsize}
\tablecaption{Predicted number of MW stars in the sample of 45 candidate members}
\tablewidth{0pt}
\tablehead{
\colhead{}&
\colhead{observed} &
\multicolumn{2}{c}{predicted interlopers} \\
\colhead{} &
\colhead{} &
\colhead{simulation} &
\colhead{scaled histogram} \\
\colhead{} &
\colhead{} &
\colhead{50\% (90\%) confidence} &
\colhead{50\% (90\%) confidence}  }
\startdata
bRGB        & 5  & $\leq$ 1 (3) & $\leq$ 1 (3) \\
fRGB        & 10 & $\leq$ 0 (2) & $\leq$ 1 (2) \\
MS/BHB     & 30 & $\leq$ 0 (2) & $\leq$ 0 (2)
\enddata
\tablecomments{The two methods for predicting the Milky Way
  contamination are described in \S3.2. It was calculated both with a
  simulation based on the Besancon Galaxy model and by scaling the
  histogram of the number of stars at velocities inconsistent with
  membership.}
\end{deluxetable} 

\begin{deluxetable}{lcccccccc}
\tabletypesize{\scriptsize}
\tablecaption{Additional data for bright candidate Willman 1 red giants}
%\tablewidth{0pt}
\tablehead{
\colhead{ID} &
\colhead{Other ID\tablenotemark{a}} &
\colhead{$u_0$\tablenotemark{b}} &
\colhead{$g_0$} &
\colhead{$r_0$} &
\colhead{$i_0$} &
\colhead{$z_0$} &
\colhead{$v_{\rm r, Martin}$ \kms\tablenotemark{c}} &
\colhead{[Fe/H]\tablenotemark{d}}
%\colhead{$\sigma_{[Fe/H]_{CaT}}$}
}
\startdata
1 & \nodata & $19.99\pm0.04$ & $18.66\pm0.02$ & $18.12\pm0.02$ & $17.87\pm0.02$ & $17.72\pm0.02$ & -- & -1.73 $\pm$ 0.12\\
2 & 1578    & $19.91\pm0.05$ & $18.69\pm0.02$ & $18.13\pm0.02$ & $17.85\pm0.02$ & $17.74\pm0.02$ & -22.0 $\pm$ 0.6 & -2.65 $\pm$ 0.12\\
3 & 1496    & $20.53\pm0.06$ & $18.81\pm0.02$ & $18.16\pm0.02$ & $17.92\pm0.02$ & $17.80\pm0.02$ & -13.2 $\pm$ 1.0 & --\\
4 & 1269    & $20.52\pm0.05$ & $18.84\pm0.02$ & $18.19\pm0.02$ & $17.98\pm0.02$ & $17.90\pm0.02$ & -10.2 $\pm$ 1.2 & --\\
6 & \nodata & $20.65\pm0.06$ & $19.39\pm0.03$ & $19.02\pm0.02$ & $18.87\pm0.02$ & $18.85\pm0.04$ & -- & --\\
\enddata
\tablecomments{Stars 1 and 6 were neither in Siegel et al.~(2008) nor Martin et al.~(2007).}
\tablenotetext{a}{ID from Siegel et al.~(2008).}
\tablenotetext{b}{These magnitudes are all extinction corrected PSF
  magnitudes from SDSS Data Release 7 \citep{dr7} so differ slightly
  from the magnitudes quoted in Table 2.}
\tablenotetext{c}{The velocities measured by \citet{martin07a}}
\tablenotetext{d}{Derived using the technique described in KGS08.}
\end{deluxetable}

\clearpage

%% If you are not including electonic art with your submission, you may
%% mark up your captions using the \figcaption command. See the 
%% No more than seven \figcaption commands are allowed per page, 
%% so if you have more than seven captions, insert a \clearpage 

\end{document}